\documentclass[letterpaper]{article} % DO NOT CHANGE THIS
\usepackage[preprint]{aaai2027}
\usepackage[hyphens]{url} % DO NOT CHANGE THIS
\usepackage{graphicx} % DO NOT CHANGE THIS
\usepackage{natbib} % DO NOT CHANGE THIS AND DO NOT ADD OPTIONS
\usepackage{caption} % DO NOT CHANGE THIS AND DO NOT ADD OPTIONS
\usepackage{booktabs}
\usepackage{amsmath}
\usepackage{amssymb}
\usepackage{array}
\usepackage{tabularx}
\usepackage{float}
\usepackage{xcolor}
\usepackage{microtype}

\newcolumntype{L}[1]{>{\raggedright\arraybackslash}p{#1}}
\newcolumntype{C}[1]{>{\centering\arraybackslash}p{#1}}

\title{
Audio-Anchored Fusion of Multi-Ratio DiT Reconstruction Residuals
for Cross-Domain Audio Deepfake Detection
}

\author{
Haotian Mo\textsuperscript{1},
Jie Liu\textsuperscript{1},
Siqi Shen\textsuperscript{2},
Songzhu Mei\textsuperscript{1},\\
Xinhai Chen\textsuperscript{1},
Xiangyang Wang\textsuperscript{1},
Yigui Feng\textsuperscript{1},
Shuai Li\textsuperscript{1},\\
Gencheng Liu\textsuperscript{1},
Keqi Yang\textsuperscript{3},
Qinglin Wang\textsuperscript{1,*}
}

\affiliations{
\textsuperscript{1}
College of Computer Science and Technology,
National University of Defense Technology,
Changsha, China\\
\textsuperscript{2}
School of Informatics,
Xiamen University,
Xiamen, China\\
\textsuperscript{3}
Hunan Zhongke Youxin Technology Co., Ltd.,
Changsha, China\\
\textsuperscript{*}
Corresponding author:
\texttt{wangqinglin@nudt.edu.cn}\\
First author:
\texttt{mohaotian25@nudt.edu.cn}
}

\begin{document}
\maketitle

\begin{abstract}
Audio deepfake detectors often degrade when generators, corpora, or recording conditions change. We use a Diffusion Transformer (DiT), trained only on bona fide speech, as a frozen reconstruction probe. Reconstructions at masking ratios 0.5, 0.75, and 0.9 yield explicit multi-ratio residual maps. Because these residuals are domain sensitive, our audio-anchored detector passes the projected frozen-WavLM auditory representation into the fusion sum without gate-based attenuation and uses residuals only as a scalar-gated additive correction. The pre-specified seed-42 run obtains 6.5442\% EER / 0.18456 min-DCF on ASVspoof~5 Eval and 13.8372\% / 0.36921 on ITW Full; three-seed means are $6.8885\pm0.3308\%$ and $15.3328\pm2.0719\%$. The latter is below a separately optimized WavLM--ResNet18 reference under both supervision settings. Auxiliary supervision raises dynamic competitive fusion from 18.4007\% to 25.2968\% mean ITW EER, worsening all three seeds. The results support reconstruction residuals as complementary evidence and motivate a non-competitive auditory path for ASVspoof~5-to-ITW transfer, without claiming a componentwise causal ablation of anchoring alone.
\end{abstract}

\section{Introduction}

High-quality text-to-speech, voice cloning, and voice conversion systems continue to lower the cost of generating realistic synthetic speech. End-to-end, diffusion-based, and flow-matching systems such as VITS, DiffWave, Grad-TTS, and F5-TTS have improved naturalness and speaker similarity \cite{kim2021vits,kong2020diffwave,popov2021gradtts,chen2025f5tts}, while increasing the risks of impersonation, telecommunication fraud, fabricated media, and attacks on voice authentication. ASVspoof 5 reflects this open setting through crowdsourced bona fide speech, diverse synthesis systems, compression, and adversarial post-processing \cite{wang2024asvspoof5scale,wang2026asvspoof5evaluation}.

Existing detectors either learn directly from waveforms or time-frequency inputs, as in RawNet2 and AASIST \cite{tak2021rawnet2,jung2022aasist}, or use large self-supervised speech encoders such as WavLM, wav2vec 2.0, and HuBERT \cite{chen2022wavlm,baevski2020wav2vec2,hsu2021hubert}. Their performance can deteriorate when generators, corpora, speakers, channels, codecs, or post-processing pipelines change \cite{muller2022generalize,liu2023asvspoof2021,pascu2024generalisable,jung2025spoofceleb}. This degradation cannot be explained only by newer attacks being intrinsically harder: distribution differences among generation and collection pipelines can invalidate correlations learned by a discriminative boundary \cite{muller2022generalize,li2024crossdomain}.

A complementary strategy is to learn a prior over bona fide data and measure which components of an input cannot be explained by that prior. Image-forensic studies have found systematic differences in frequency statistics and diffusion denoising trajectories \cite{frank2020frequency,liang2025trajectory}. In audio, masked autoencoding and genuine-focused reconstruction have also been used to expose synthetic-speech abnormalities \cite{huang2022audiomae,wang2024gflfad}. However, it remains unclear how to convert a bona-fide-trained DiT into explicit forensic evidence and how to combine that evidence with a strong auditory representation without amplifying cross-domain errors.

We train a conditional-flow-matching DiT exclusively on bona fide speech. At detection time, the frozen probe reconstructs masked Mel spectrograms, and the absolute input--reconstruction difference is retained as an explicit residual. Unlike a scalar reconstruction error, this representation preserves the time location, frequency region, and relative strength of each mismatch. Because the probe never observes spoof labels or attack identities, its residual is not directly optimized to separate training attacks.

\begin{figure}[t]
    \centering
    \includegraphics[width=\columnwidth]{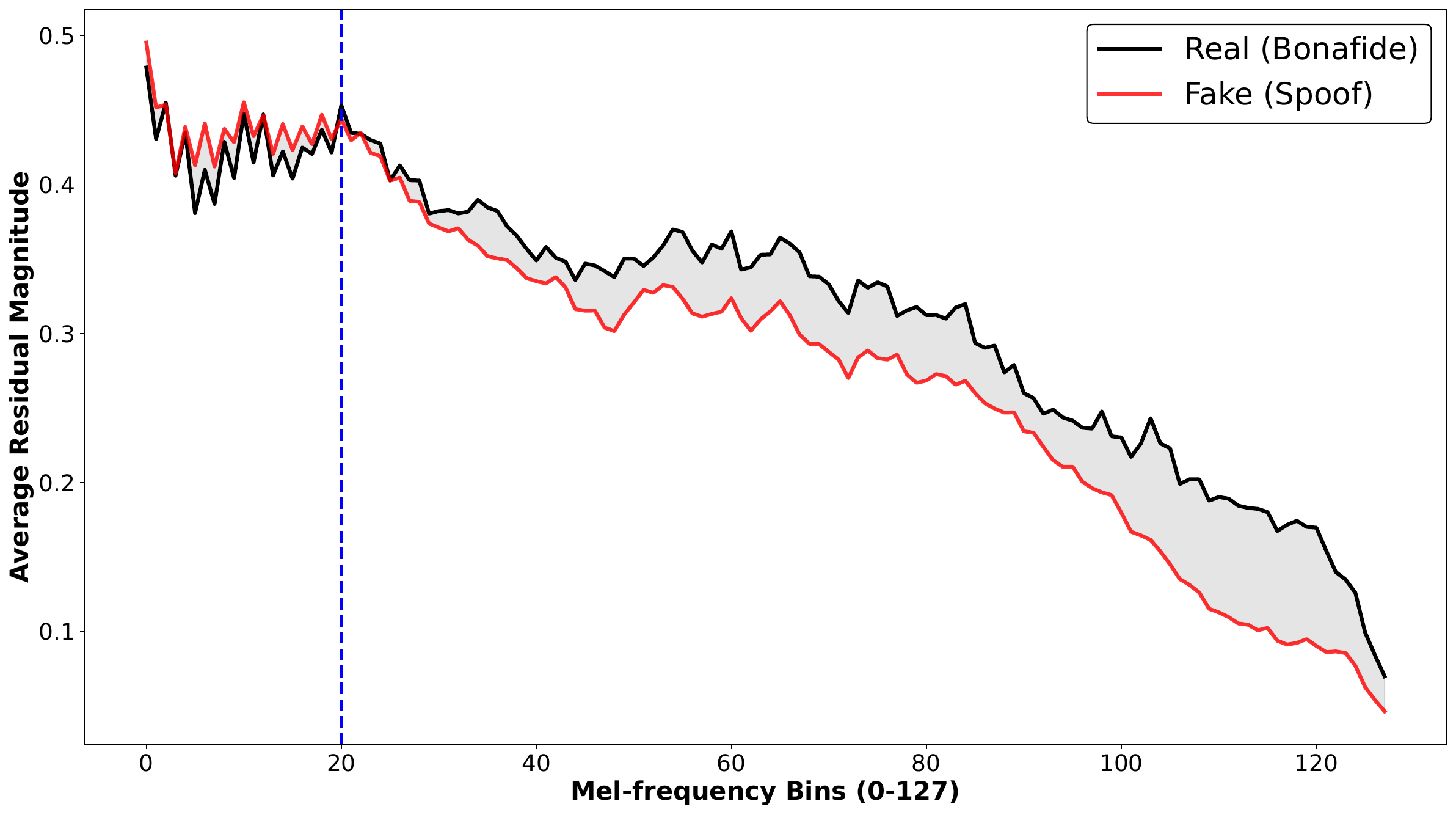}
    \caption{Mean DiT reconstruction residual spectra on the balanced ASVspoof~5 diagnostic probe spanning A01--A32 at masking ratio 0.9. Residuals are averaged over time per utterance and then over samples, without sample-wise normalization. The dashed line is a visual reference only, not a predefined boundary.}
    \label{fig:residual-spectrum}
\end{figure}

Figure~\ref{fig:residual-spectrum} provides an initial diagnostic. At masking ratio 0.9, the mean residual spectra of bona fide and spoofed speech do not differ through a uniform amplitude offset; their relative ordering changes across frequency regions, approximately near Mel bin 20. This location is not imposed as a fixed boundary. The observation instead indicates that the residual is structured and may retain local components that are difficult for a bona fide reconstruction prior to explain.

Structured residuals are not automatically domain robust. Their magnitude and frequency distribution may depend on linguistic content, speaker, channel, codec, and reconstruction difficulty. Masking ratio adds another source of variation: lower ratios preserve more local context, whereas higher ratios force the probe to rely more heavily on the bona fide prior. A single ratio can therefore expose only part of the reconstruction mismatch.

Residual domain sensitivity also makes fusion consequential. WavLM provides a broad auditory representation, whereas DiT residuals describe local deviations from a reconstruction prior. In competitive fusion, increasing the residual weight simultaneously suppresses the auditory weight. Under a target-domain residual shift, the model may both introduce unreliable evidence and weaken an otherwise useful auditory representation. We instead constrain residuals to act as a correction: the auditory branch enters the shared pre-projection fusion sum directly, while the gate scales only the residual term.

Our contributions are threefold:
\begin{itemize}
    \item We construct explicit multi-ratio DiT reconstruction residuals as active forensic evidence. A large-scale bona-fide-trained probe reconstructs inputs at three masking ratios, preserving localized mismatches rather than collapsing them into a scalar error.
    \item We propose audio-anchored additive fusion. The WavLM representation enters the pre-projection fusion sum without gate-based scaling, while a sample-level scalar gate controls only the residual correction and cannot explicitly down-weight the auditory branch.
    \item We evaluate against a separately optimized WavLM--ResNet18 single-stream reference and report three-seed system-level comparisons between two residual-based designs under shared data, cache, auditory branch, objectives, and optimizer. The strongest repeated observation is that auxiliary supervision harms dynamic competitive fusion on ITW for all three seeds.
\end{itemize}

\section{Related Work}

\paragraph{Audio Deepfake Detection and Self-Supervised Speech Models.}
Early systems relied on handcrafted cepstral features and shallow classifiers. End-to-end models such as RawNet2 and AASIST instead learn waveform or spectro-temporal artifacts directly \cite{tak2021rawnet2,jung2022aasist}. Self-supervised models further improve representation quality through WavLM, wav2vec 2.0, and HuBERT front ends \cite{chen2022wavlm,baevski2020wav2vec2,hsu2021hubert}. Multi-view and expert-fusion systems combine heterogeneous evidence \cite{zhang2025multiview,wang2024moe}, and audio large language models have recently been explored for deepfake detection \cite{gu2025allm4add}. Genuine-oriented systems such as SLIM further improve transfer by learning speech dependencies from bona fide data, although their final detector training may also benefit from augmentation and longer or full-length inputs \cite{zhu2024slim}. In contrast, we retain explicit reconstruction residual maps from an independently trained bona-fide generative probe and study how such domain-sensitive evidence should be constrained during fusion. Nevertheless, cross-generator and cross-corpus generalization remains difficult \cite{muller2022generalize,pascu2024generalisable,li2024crossdomain}.

\paragraph{Reconstruction-Based Detection and Generative Priors.}
Reconstruction-based anomaly detection learns regularities from normal data and identifies inputs that are poorly explained by the learned prior. Masked autoencoders provide a practical mechanism for learning such structure \cite{he2022mae,huang2022audiomae}. Genuine-focused reconstruction and one-class anti-spoofing reduce dependence on known spoof classes \cite{wang2024gflfad,zhang2021oneclass}. Recent work has also used diffusion reconstruction to synthesize hard examples for generalizable detection \cite{cheng2026diffrecon} and diffusion models to expose time--frequency artifact regions for explanation \cite{grinberg2025diffexplain}. We do not claim the first use of bona-fide-only or diffusion-based reconstruction. Our focus is on caching explicit multi-ratio residual maps from a large-scale bona-fide probe and studying their asymmetric fusion with a strong auditory representation.

\paragraph{Heterogeneous Evidence Fusion.}
Frozen-SSL expert fusion can combine pretrained representations through sample-adaptive weights \cite{wang2024moe}, but higher fusion freedom does not guarantee robust transfer. Competitive weighting may overuse a shifted branch and suppress a more stable one. Our asymmetric rule instead leaves the auditory branch outside the gate and permits reconstruction evidence only as an additive pre-projection correction.

\section{Method}

\begin{figure*}[t]
    \centering
    \includegraphics[width=0.95\textwidth]{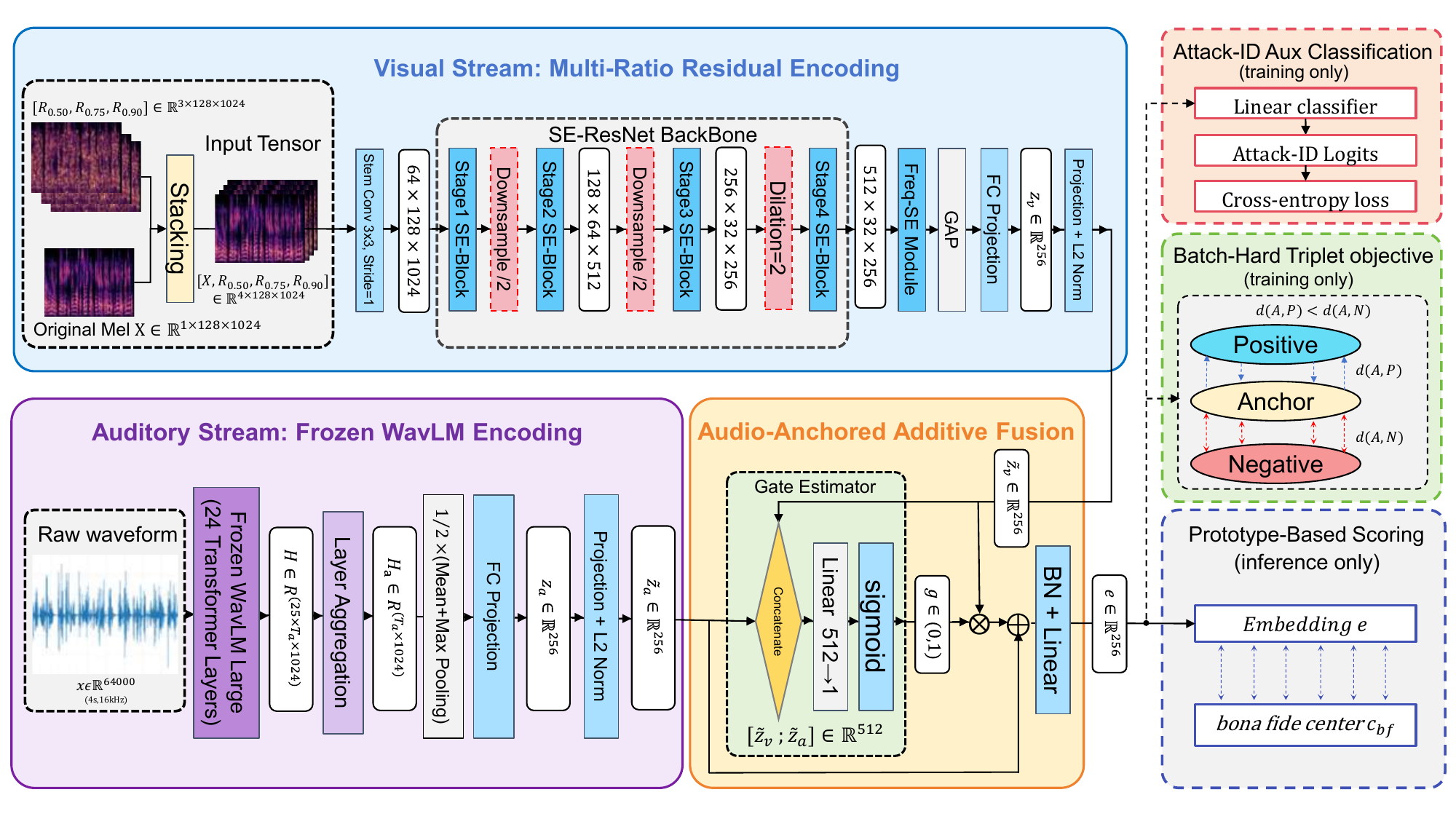}
    \caption{Audio-anchored dual-stream detector. The scalar-gated spectrogram--residual representation is added to the frozen WavLM-Large anchor. The final embedding $e$ feeds the training-only Attack-ID and triplet objectives; prototype scoring is inference only.}
    \label{fig:overview}
\end{figure*}

Figure~\ref{fig:overview} summarizes the framework. A bona-fide-trained DiT generates multi-ratio spectrogram residuals offline. The spectrogram--residual stream encodes the original Mel spectrogram and three residual maps, while a frozen WavLM-Large stream encodes the waveform. Audio-anchored fusion passes the auditory vector directly into the fusion sum and uses a sample-level scalar gate only for the additive residual correction. After batch normalization and the shared linear projection, the final embedding $e$ feeds both training objectives; at inference, the same embedding is scored against a fixed bona fide prototype constructed from ASVspoof 5 Train.

\subsection{Bona Fide DiT Reconstruction Probe}

Following the conditional flow-matching DiT backbone of F5-TTS \cite{chen2025f5tts}, we train a masked reconstruction probe using only bona fide speech. Let $X\in\mathbb{R}^{F\times T}$ be a Mel spectrogram and $M^{(r)}\in\{0,1\}^{F\times T}$ a binary mask with ratio $r$. The visible condition is
\begin{equation}
X_{\mathrm{vis}}^{(r)}=(1-M^{(r)})\odot X.
\end{equation}
A frozen Audio-MAE encoder provides a global condition $c$. Conditional flow matching interpolates between Gaussian noise $X_0$ and the target spectrogram:
\begin{equation}
X_t=(1-t)X_0+tX, \qquad u_t=X-X_0.
\end{equation}
The DiT predicts the masked-region velocity field,
\begin{equation}
\mathcal{L}_{\mathrm{DiT}}=
\mathbb{E}\left[\left\|M^{(r)}\odot\left(v_\theta(X_t,t,X_{\mathrm{vis}}^{(r)},c)-u_t\right)\right\|_2^2\right].
\end{equation}
After training, both modules are frozen. Independent Bernoulli masks have expected ratio $r$; visible cells are copied from $X$ and held at zero velocity, so Equation~(4) measures masked-region mismatch. One seeded 16-step Euler reconstruction per utterance and ratio is cached as $\widehat{X}^{(r)}$.

\begin{figure}[t]
    \centering
    \includegraphics[width=0.86\columnwidth]{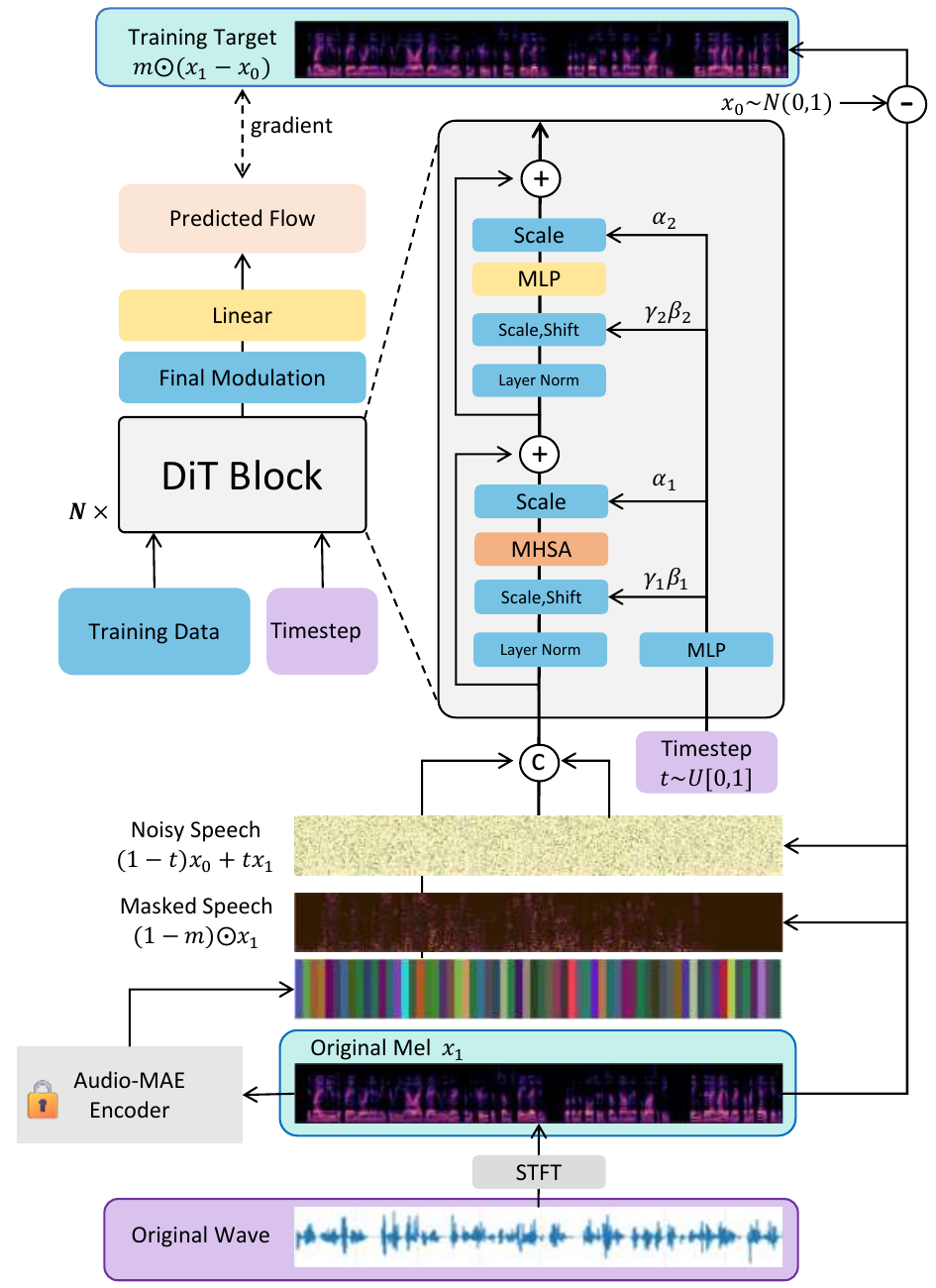}
    \caption{Masked self-supervised training of the DiT reconstruction probe. Only bona fide speech is used. The probe predicts conditional flow in masked spectrogram regions from the visible spectrogram and the frozen Audio-MAE condition.}
    \label{fig:dit-training}
\end{figure}

\subsection{Multi-Ratio Spectrogram--Residual Encoding}

We use masking ratios $\mathcal{R}=\{0.5,0.75,0.9\}$ and compute
\begin{equation}
R^{(r)}=\left|X-\widehat{X}^{(r)}\right|, \qquad r\in\mathcal{R}.
\end{equation}
The stream input is the channel-wise stack
\begin{equation}
V=\operatorname{Concat}\left[X,R^{(0.5)},R^{(0.75)},R^{(0.9)}\right].
\end{equation}
We do not apply sample-wise InstanceNorm at the input because absolute residual magnitude, cross-ratio energy, and frequency-dependent statistics may carry forensic information. The encoder is a modified ResNet-18 with a small-stride stem, no initial max pooling, squeeze-and-excitation blocks, dilated convolutions, and a lightweight frequency squeeze-and-excitation module. Global pooling and projection produce a 256-dimensional residual representation $z_v$.

\subsection{Frozen WavLM Auditory Encoding}

Let $H_a^{(0)},\ldots,H_a^{(L)}$ denote the convolutional and Transformer-layer states of frozen WavLM-Large. Learnable normalized weights aggregate the layers:
\begin{equation}
H_a=\sum_{l=0}^{L}\alpha_l H_a^{(l)},\qquad
\alpha_l=\frac{\exp(w_l)}{\sum_{j=0}^{L}\exp(w_j)}.
\end{equation}
Mean and max temporal pooling are averaged and projected to obtain the 256-dimensional auditory representation $z_a$. Only the layer weights and projection are updated; the WavLM backbone remains frozen.

\subsection{Audio-Anchored Additive Fusion}

Let $W_a$ and $W_v$ be learned projections into a shared 256-dimensional space. After $L_2$ normalization, a sample-level scalar gate is estimated from the concatenated branch representations:
\begin{equation}
\begin{aligned}
\widetilde{z}_a&=\frac{W_a z_a}{\|W_a z_a\|_2},&
\widetilde{z}_v&=\frac{W_v z_v}{\|W_v z_v\|_2},\\
g&=\sigma\!\left(w_g^{\top}[\widetilde{z}_v;\widetilde{z}_a]+b_g\right),&
g&\in(0,1).
\end{aligned}
\end{equation}
The pre-fusion representation is
\begin{equation}
z_f=\widetilde{z}_a+g\widetilde{z}_v.
\end{equation}
Thus, the auditory vector enters the pre-projection sum with coefficient one, while the scalar gate controls only the residual correction. Batch normalization and a shared linear projection produce $e\in\mathbb{R}^{256}$. Anchoring guarantees that the residual gate cannot explicitly suppress the auditory branch, not that the final embedding is invariant to residual evidence.

For comparison, the dynamic competitive system introduces sample-dependent routing at both the residual-input and modality levels. Let $m_r=\mu(R^{(r)})$ and $a_r=\mu(|R^{(r)}|)$ denote the mean and mean absolute residual at ratio $r$. The router descriptor is
\begin{equation}
q=[m_{0.5},m_{0.75},m_{0.9},a_{0.5},a_{0.75},a_{0.9}].
\end{equation}
The router predicts $\pi=\operatorname{softmax}(f_r(q))$ and rescales each residual channel before visual encoding, $\bar R^{(r)}=\pi_r R^{(r)}$. Separate Mel and residual stems merge after the first stage to produce $z_v^{\mathrm{dyn}}$. A scalar modality gate then applies complementary weights before concatenation and projection:
\begin{equation}
\begin{aligned}
\gamma&=\sigma\!\left(w_c^{\top}[z_v^{\mathrm{dyn}};z_a]+b_c\right),\\
e^{\mathrm{comp}}&=W_c\,\mathrm{BN}([\gamma z_v^{\mathrm{dyn}};(1-\gamma)z_a]).
\end{aligned}
\end{equation}
Increasing the residual coefficient necessarily decreases the auditory coefficient. The system shares data, cache, auditory branch, objectives, and optimizer with ours, but also changes routing and visual organization; it is a system-level contrast, not a componentwise anchoring ablation.

\subsection{Training Objectives and Prototype Scoring}

Batch-hard triplet loss organizes bona fide and spoof embeddings:
\begin{equation}
\mathcal{L}_{\mathrm{tri}}=\frac{1}{B}\sum_{i=1}^{B}
\max\left(0,m+d(e_i,e_{p(i)})-d(e_i,e_{n(i)})\right).
\end{equation}
The auxiliary head uses a fixed 33-output index for bona fide and A01--A32, but positive targets occur only for the nine observed Train labels (bona fide and A01--A08); Dev/Eval identities never enter training. The fixed indexing is an implementation convention. With label smoothing 0.1, the total objective is
\begin{equation}
\mathcal{L}=\mathcal{L}_{\mathrm{tri}}+\lambda_{\mathrm{aux}}\mathcal{L}_{\mathrm{aux}},
\qquad \lambda_{\mathrm{aux}}=0.2,
\end{equation}
with $\lambda_{\mathrm{aux}}=0$ for controls without auxiliary supervision.

At inference, no binary output head is used. We normalize all bona fide ASVspoof 5 Train embeddings and construct
\begin{equation}
c_{bf}=\frac{\sum_{i:y_i=bf}e_i/\|e_i\|_2}{\left\|\sum_{i:y_i=bf}e_i/\|e_i\|_2\right\|_2}.
\end{equation}
The spoof score is
\begin{equation}
s(x)=-\left(\frac{e(x)}{\|e(x)\|_2}\right)^{\top}c_{bf}.
\end{equation}
For the residual-based systems, the same training-set prototype is used for Dev, Eval, and ITW. All final Eval and ITW results, including the single-stream baseline, use a prototype recomputed from ASVspoof 5 Train; no target-domain center update or threshold calibration is performed. Prototype scoring is treated as a fixed, uncalibrated ranking protocol rather than an independently validated contribution.

\section{Experimental Setup}

The DiT and detector are trained in two decoupled stages. The DiT uses 2,552,125 bona fide utterances from ASVspoof 5 Train, ASVspoof 2019 Train, Common Voice, VoxCeleb2, and LibriSpeech. Table~\ref{tab:dit-training-data} gives the exact composition. Fixed-source sampling forms each 40-utterance DiT batch in a 14:2:9:7:8 ratio, respectively, so the much larger external corpora do not remove task-related bona fide conditions from training.

% Required packages are loaded in the preamble.

\begin{table}[t]
\centering
\caption{Bona fide corpora used to train the DiT reconstruction probe.
All spectrograms are standardized to 1,024 Mel frames.}
\label{tab:dit-training-data}

\scriptsize
\setlength{\tabcolsep}{3.2pt}
\renewcommand{\arraystretch}{1.08}

\begin{tabularx}{\columnwidth}{
    @{}
    >{\raggedright\arraybackslash}p{0.23\columnwidth}
    >{\raggedright\arraybackslash}X
    >{\raggedleft\arraybackslash}p{0.19\columnwidth}
    >{\raggedleft\arraybackslash}p{0.13\columnwidth}
    @{}
}
\toprule
\textbf{Source}
& \textbf{Used subset}
& \textbf{Utterances}
& \textbf{Share} \\
\midrule

ASVspoof 5
& Train bona fide
& 18,797
& 0.74\% \\

ASVspoof 2019
& Train bona fide
& 2,580
& 0.10\% \\

Common Voice
& Filtered valid speech
& 1,157,498
& 45.35\% \\

VoxCeleb2
& All valid speech
& 1,092,009
& 42.79\% \\

LibriSpeech
& train-clean-100/360/500
& 281,241
& 11.02\% \\

\midrule
\textbf{Total}
& \textbf{Bona fide only}
& \textbf{2,552,125}
& \textbf{100\%} \\
\bottomrule
\end{tabularx}
\end{table}

The detector uses only ASVspoof 5 Train attacks A01--A08. Dev attacks A09--A16 are used only for checkpoint selection; Eval attacks A17--A32 and ITW Full are read-only final evaluations. ITW Full merges its train, validation, and test partitions. Table~\ref{tab:detector-data} reports the exact sample counts, mean durations, and roles.

\paragraph{Decoupled residual generation and detector training.} After reconstruction training, both DiT and the Audio-MAE condition encoder remain frozen. We first process every split offline: each utterance is converted to a Mel spectrogram, reconstructed independently at ratios 0.5, 0.75, and 0.9, and stored together with the three absolute residual maps. Detector training never calls DiT online. The spectrogram--residual stream reads the same four-channel cache for every fusion structure and ablation, while the auditory stream reads the corresponding waveform through its own cropping protocol. This decoupling prevents stochastic reconstruction differences from confounding comparisons among input normalization, routing, fusion, and auxiliary-supervision settings.

\begin{table}[t]
\centering
\caption{Detector training and evaluation data.}
\label{tab:detector-data}

\tiny
\setlength{\tabcolsep}{2.0pt}
\renewcommand{\arraystretch}{0.96}

\begin{tabular*}{\columnwidth}{
    @{\extracolsep{\fill}}
    l
    r
    r
    r
    c
    l
    @{}
}
\toprule
\textbf{Split}
& \textbf{Bona fide}
& \textbf{Spoof}
& \textbf{Total}
& \textbf{Mean (s)}
& \textbf{Role} \\
\midrule
ASV5 Train
& 18,797
& 163,560
& 182,357
& 11.9
& Train/prototype \\

ASV5 Dev
& 31,334
& 109,616
& 140,950
& 7.1
& Checkpoint selection \\

ASV5 Eval
& 138,688
& 542,086
& 680,774
& 7.1
& Unseen attacks \\

ITW Full
& 19,963
& 11,816
& 31,779
& 4.3
& Cross-corpus test \\
\bottomrule
\end{tabular*}
\end{table}

All audio is resampled to 16 kHz. The DiT and residual stream use the leading 1,024 Mel frames (about 10.24 s at a 160-sample hop), whereas WavLM uses 4-second waveforms: random crops for training and leading crops otherwise, with padding when needed. Thus, the streams come from the same utterance but have different temporal supports. Each seeded residual realization is generated once with 16 Euler steps per ratio and reused by every residual-system run. Exact architecture, masking, solver, preprocessing, and cache-generation settings are consolidated in Appendix~\ref{app:protocol}.

The four residual-based configurations use five AdamW epochs, learning rate $10^{-3}$, weight decay $10^{-4}$, one warm-up epoch, cosine decay, and four-step gradient accumulation. Physical batches contain 32 bona fide and 32 attack-balanced spoofed samples; batch-hard mining uses margin 0.3. The 256-dimensional models use a 33-way source head with label smoothing 0.1 and weight 0.2 when enabled. Seeds 42, 123, and 2026 are selected by Dev EER using the Train prototype.

\paragraph{Strong single-stream reference.} We separately optimize WavLM--ResNet18: aggregated frozen WavLM states form a single-channel time--feature map that a randomly initialized ResNet-18 maps to 256 dimensions. It uses learning rate $10^{-4}$, binary class-balanced sampling, and its original random-crop/Dev-center selection procedure; final Eval and ITW use leading crops and the Train prototype. It is therefore a strong reference, not a one-branch ablation. Full protocol differences are reported in Appendix~\ref{app:protocol}. We report three-seed mean and sample standard deviation; EER and normalized min-DCF follow ASVspoof 5 \cite{wang2024asvspoof5scale}.

We use ASVspoof 5 as a post-challenge research benchmark. Because WavLM-Large and external bona-fide DiT pretraining fall outside official constraints, we make no open/closed, eligibility, or leaderboard claim. Public systems and WavLM--ResNet18 provide context. The residual-system comparison shares the principal training protocol but is not componentwise matched, so it supports a structural interpretation rather than isolating anchoring as the sole causal factor.

\section{Results and Analysis}

\subsection{Main Results and Comparison with Public Systems}

Table~\ref{tab:main-results} separates unmatched public results, a separately optimized strong single-stream reference, and residual systems trained under the shared protocol. Because augmentation, data, duration, scoring, and optimization affect generalization \cite{sun2024automated}, the public rows are contextual and should not be read as a ranking. Wav2Vec2-AASIST uses augmentation and extra ASVspoof 2019 LA supervision; SLIM uses RawBoost, longer training segments, and full-length evaluation. WavLM--ResNet18 has its own backend and optimization, whereas the four residual systems share the auditory branch, cache, losses, and optimizer. External bona fide data train only the frozen DiT.

Seed 42 was fixed before repeated-seed validation. Against WavLM--ResNet18, audio anchoring changes Eval/ITW EER from 7.2699\%/27.1452\% to 6.6809\%/19.2282\% without Aux. With Aux, Eval is similar (6.5442\% versus 6.5226\%), while ITW improves from 18.5994\% to 13.8372\%. This establishes competitiveness against a strong single-stream reference, not a one-variable residual ablation; fusion attribution uses the matched residual systems.

Dynamic residual routing with competitive fusion obtains the lowest internal primary-run Eval EER, 6.0208\%, but degrades to 20.9337\% on ITW. The repeated-seed results in Table~\ref{tab:aux-interaction} and Appendix~\ref{app:three-seed} evaluate whether these structural trends persist beyond the primary run.

\begin{table}[!b]
\centering
\caption{Contextual public results, the separately optimized single-stream reference, and pre-specified seed-42 residual-system runs. Public rows are not directly comparable; lower is better.}
\label{tab:main-results}
\tiny
\setlength{\tabcolsep}{2.0pt}
\renewcommand{\arraystretch}{0.94}
\begin{tabularx}{\columnwidth}{L{0.44\columnwidth} *{4}{>{\centering\arraybackslash}X}}
\toprule
& \multicolumn{2}{c}{ASVspoof 5 Eval} & \multicolumn{2}{c}{ITW Full} \\
\cmidrule(lr){2-3} \cmidrule(lr){4-5}
Method & EER & min-DCF & EER & min-DCF \\
\midrule
\multicolumn{5}{l}{\textit{Representative public systems}} \\
RawNet2 official baseline \cite{wang2024asvspoof5scale} & 36.04 & 0.8266 & -- & -- \\
AASIST official baseline \cite{wang2024asvspoof5scale} & 29.12 & 0.7106 & -- & -- \\
Fused SSL + Improved NeXt-TDNN \cite{tahaoglu2025nexttdnn} & 7.23 & -- & -- & -- \\
Wav2Vec2-AASIST$^{\dagger}$ \cite{schafer2024robust} & 6.06 & 0.174 & -- & -- \\
SLIM$^{\dagger}$ \cite{zhu2024slim} & 5.56 & 0.1499 & 10.8 & -- \\
\midrule
\multicolumn{5}{l}{\textit{Strong single-stream reference}} \\
WavLM--ResNet18 (without source auxiliary loss) & 7.2699 & 0.20943 & 27.1452 & 0.61732 \\
WavLM--ResNet18 (with source auxiliary loss) & 6.5226 & 0.17533 & 18.5994 & 0.40741 \\
\midrule
\multicolumn{5}{l}{\textit{Residual systems under the shared protocol}} \\
Audio-anchored additive fusion (without source auxiliary loss) & 6.6809 & 0.18702 & 19.2282 & 0.50756 \\
Dynamic residual routing with competitive fusion (without source auxiliary loss) & \textbf{6.0208} & \textbf{0.16414} & 20.9337 & 0.47368 \\
\textbf{Audio-anchored additive fusion (complete model)} & 6.5442 & 0.18456 & \textbf{13.8372} & \textbf{0.36921} \\
\bottomrule
\end{tabularx}
\vspace{1pt}
\parbox{0.98\columnwidth}{\tiny $^{\dagger}$Supervised detector training uses augmentation. Wav2Vec2-AASIST also uses ASVspoof 2019 LA; SLIM uses full-length validation/inference and reports RawBoost reducing ITW EER from 25.7\% to 10.8\%. WavLM--ResNet18 is separately optimized; the residual rows form the controlled fusion comparison. Seed 42 was fixed before repeated validation. Three-seed results are in Table~\ref{tab:aux-interaction} and Appendix~\ref{app:three-seed}. These post-challenge systems are not assigned official open/closed status.}
\end{table}

\subsection{Multi-Ratio Reconstruction Residuals}

\begin{table}[t]
\centering
\caption{Detection performance with single-ratio and multi-ratio reconstruction residuals.}
\label{tab:ratio-ablation}

\scriptsize
\setlength{\tabcolsep}{2.6pt}
\renewcommand{\arraystretch}{0.96}

\begin{tabular*}{\columnwidth}{
    @{\extracolsep{\fill}}
    l
    r
    r
    r
    @{}
}
\toprule
\textbf{Residual input}
& \textbf{Dev EER (\%)}
& \textbf{Eval EER (\%)}
& \textbf{Eval min-DCF} \\
\midrule
Masking ratio 0.5
& 1.3911
& 7.1578
& 0.19922 \\

Masking ratio 0.75
& 1.3648
& 7.5128
& 0.21005 \\

Masking ratio 0.9
& 2.1248
& 6.9716
& 0.19599 \\

Ratios 0.5 + 0.75 + 0.9
& 1.8586
& \textbf{6.5442}
& \textbf{0.18456} \\
\bottomrule
\end{tabular*}
\end{table}

Table~\ref{tab:ratio-ablation} shows that each single ratio can obtain a low Dev EER, and ratio 0.75 even outperforms the multi-ratio input on Dev. The ranking reverses on A17--A32 Eval: every single-ratio system degrades, and the strongest single ratio, 0.9, remains 0.4274 EER points worse than the three-ratio input. A reconstruction difficulty selected on a limited set of development attacks therefore does not stably cover new generation mechanisms.

\subsection{Raw Residual Statistics and Fusion Structure}

Sample-wise InstanceNorm consistently degrades Eval performance across four paired structures. Raw residual inputs average 1.3700\% Dev EER, 6.8422\% Eval EER, and 0.19071 min-DCF, whereas InstanceNorm averages 1.8200\%, 7.3484\%, and 0.20549. This supports normalization inside the network while retaining the absolute residual magnitude and cross-ratio energy relations at the input.

The residual systems share data, cache, auditory branch, objectives, optimizer, and scoring, but differ in routing, visual organization, and fusion geometry. Their gap is therefore system level, not an isolated anchoring ablation. Dynamic competitive fusion has worse mean ITW EER under both supervision settings; with Aux, audio anchoring is better in all three paired seeds, consistent with preserving the auditory coefficient.

\subsection{Interaction with Source Auxiliary Supervision}

\begin{table}[t]
\centering
\caption{Three-seed EER (\%). WavLM--ResNet18 is separately optimized; the residual-based rows share the controlled protocol. Values are mean$\pm$sample standard deviation.}
\label{tab:aux-interaction}

\tiny
\setlength{\tabcolsep}{1.8pt}
\renewcommand{\arraystretch}{0.94}

\begin{tabular*}{\columnwidth}{
    @{\extracolsep{\fill}}
    l
    c
    c
    c
    c
    @{}
}
\toprule
\textbf{System}
& \textbf{Aux.}
& \textbf{Dev}
& \textbf{ASV5 Eval}
& \textbf{ITW Full} \\
\midrule
WavLM--ResNet18 & No & $4.8707\pm0.4186$ & $6.9306\pm0.3420$ & $22.5704\pm4.7675$ \\
WavLM--ResNet18 & Yes & $4.9445\pm0.2190$ & $7.0603\pm0.5564$ & $18.2738\pm4.9847$ \\
Audio-anchored & No & $2.4100\pm1.0732$ & $7.1112\pm0.3727$ & $17.1328\pm5.8921$ \\
Audio-anchored & Yes & $1.7595\pm0.3997$ & $6.8885\pm0.3308$ & $\mathbf{15.3328\pm2.0719}$ \\
Dynamic competitive & No & $2.9382\pm0.3183$ & $6.7901\pm0.8929$ & $18.4007\pm2.5810$ \\
Dynamic competitive & Yes & $1.5041\pm0.3323$ & $7.1585\pm0.2819$ & $25.2968\pm1.8357$ \\
\bottomrule
\end{tabular*}
\end{table}

Relative to WavLM--ResNet18, audio anchoring has lower paired ITW EER in two of three seeds under either supervision setting; we therefore claim a lower mean, not initialization-independent superiority. With Aux, it beats dynamic competitive fusion in all seeds by 9.37, 12.17, and 8.36 points, while Aux degrades that system in every seed. With $n=3$, we make no significance claim; the pattern supports a fusion--supervision interaction, not causality from anchoring alone.

\subsection{Embedding-Space Visualization}

\begin{figure}[t]
    \centering
    \includegraphics[width=0.78\columnwidth]{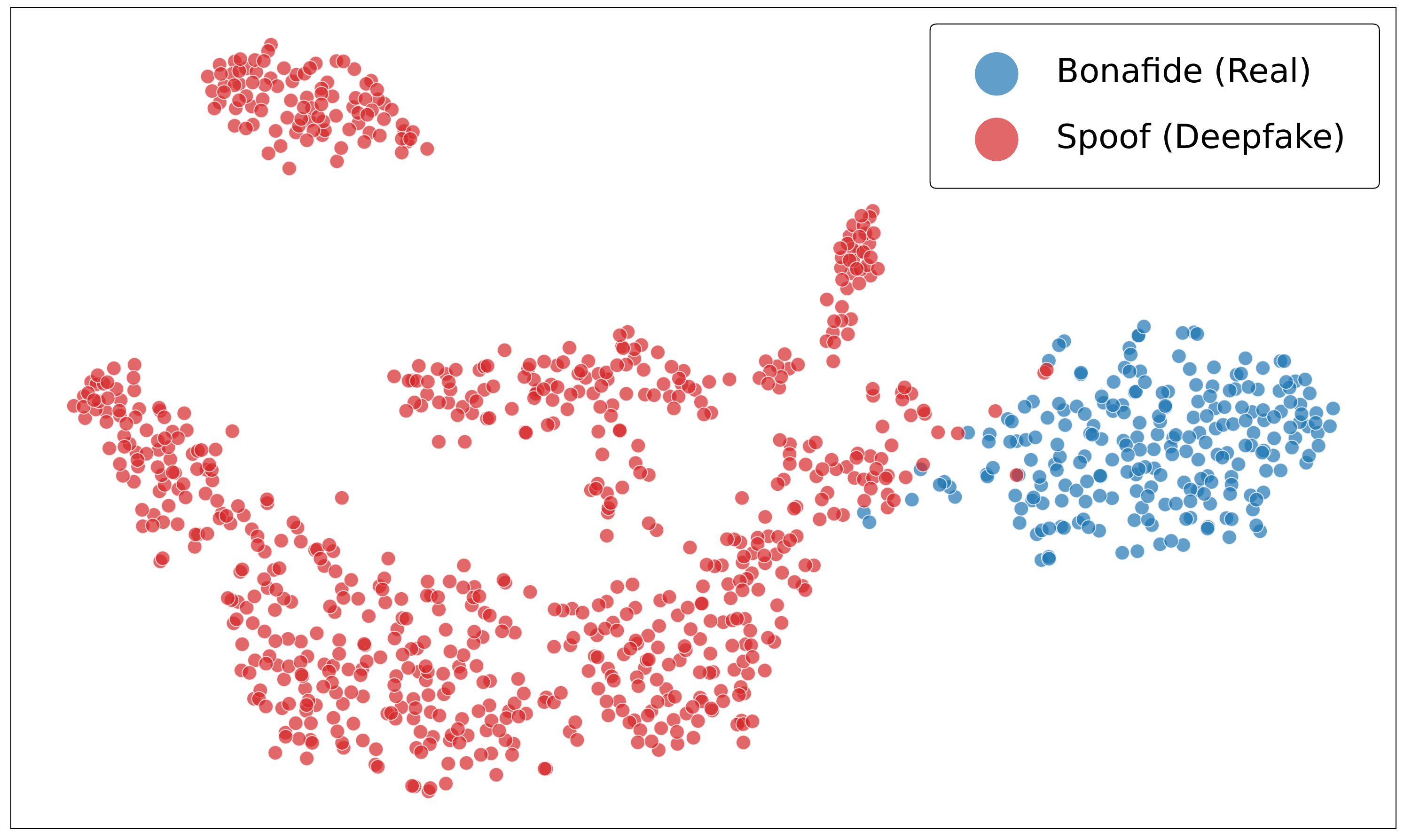}
    \caption{Illustrative t-SNE of 1,000 ASVspoof~5 Dev embeddings. The plot is qualitative only and is not used for model selection, statistical inference, or cross-corpus evidence.}
    \label{fig:tsne}
\end{figure}

Because t-SNE distorts global distances and uses Dev data, Figure~\ref{fig:tsne} is illustrative only.

\subsection{Post-Hoc Per-Attack Subband Diagnosis}

After all model choices and primary experiments were completed, we used the same balanced diagnostic probe as Figure~\ref{fig:residual-spectrum} for a post-hoc analysis over A01--A32. Residuals were first averaged over time, and the 128 Mel bins were divided into eight equal subbands. For each attack, the mean residual within each subband at ratio 0.9 was used as a scalar score and its EER was computed independently. This analysis did not participate in training, ratio selection, fusion design, checkpoint selection, or calibration.

\begin{figure}[t]
    \centering
    \includegraphics[width=0.94\columnwidth]{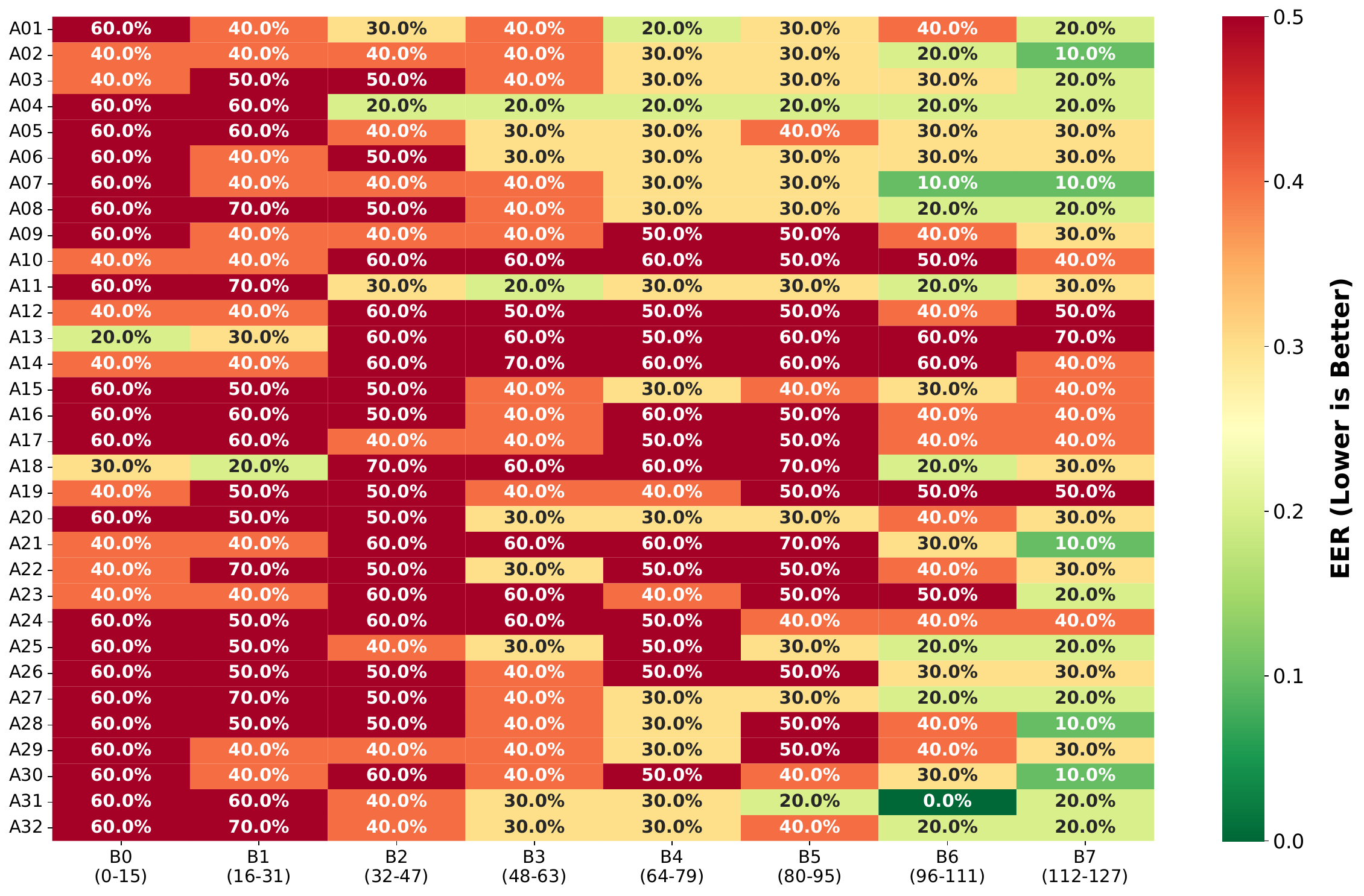}
    \caption{Post-hoc per-attack subband EER diagnostic for A01--A32. The most discriminative subband varies across attacks, and no fixed high-frequency boundary is consistently optimal. The diagnostic is explanatory only and is not used for model or threshold selection.}
    \label{fig:subband}
\end{figure}

Figure~\ref{fig:subband} shows substantial frequency heterogeneity across generation mechanisms. The preferred subband changes from attack to attack, so the trend change in Figure~\ref{fig:residual-spectrum} should not be interpreted as a universal boundary. This observation is consistent with the advantage of retaining fine-grained multi-ratio residual maps, but it does not by itself establish cross-corpus generalization.

\subsection{Computational Cost}

DiT reconstruction dominates cost: three residual maps require 2.37 s per utterance, versus 24.4 ms for cached detection, and about 1.00 MiB storage. For 100,000 utterances, linear serial scaling is about 65.8 device-hours and 97.7 GiB before parallelization, positioning the system for offline review rather than streaming.

\section{Discussion}

\paragraph{What the comparisons establish.} WavLM--ResNet18 is a strong reference, not a pooled weak baseline, so the lower mean ITW error supports practical complementarity. Its backend, optimization, and temporal support differ, however, and the residual systems also differ in routing and visual organization. We therefore make neither a one-variable anchoring claim nor a matched-data DiT-superiority claim; the 2.55-million-utterance probe is evaluated as a fixed system component.

\paragraph{Why anchoring is a structural safeguard.} The scalar gate does not detect domain shift or guarantee output invariance; it only cannot explicitly reduce the auditory coefficient before projection. Aux improves Dev for dynamic competitive fusion but raises mean ITW EER from $18.4007\%$ to $25.2968\%$ in all seeds, consistent with this motivation. Because the contrast also changes routing and visual organization, competitive suppression is not claimed as the sole cause.

\paragraph{Training-head and inference consistency.} Figure~\ref{fig:overview} makes the shared endpoint explicit: after batch normalization and the shared linear projection, $e$ feeds both training heads. The Attack-ID head is discarded at inference; triplet mining also acts on $e$, and testing uses the same embedding with the fixed ASVspoof~5 Train prototype, without auxiliary logits or target-domain calibration. Thus, auxiliary effects reflect representation learning rather than scoring-rule changes. All residual systems reuse the same frozen DiT cache, and their fixed 1,024-frame residual and 4-second auditory supports are not frame-synchronous.

\paragraph{Repeated-seed robustness.} Seed 42 was fixed before repeated validation. Audio anchoring with Aux beats WavLM--ResNet18 in seeds 42 and 123 but not 2026, so the lower mean ($15.3328\pm2.0719\%$ versus $18.2738\pm4.9847\%$) is not an initialization-independent or significant claim. It beats dynamic competitive fusion with Aux in all paired seeds; Aux also worsens that system on ITW in every seed despite better Dev EER.

\paragraph{Auxiliary-label scope.} The 33-output head preserves a common index for bona fide and A01--A32, but only bona fide and A01--A08 are positive training targets. The unused outputs remain non-target classes in the softmax denominator; they do not expose Dev or Eval identities to training. The observed fusion--supervision interaction is specific to label smoothing 0.1 and $\lambda_{\mathrm{aux}}=0.2$. Without a weight sweep, it should not be generalized to every form or strength of source supervision.

\paragraph{Residual interpretation.} The probe uses independent Bernoulli masks, copies visible Mel cells from the input, and holds their flow velocity at zero. Consequently, visible entries contribute zero to the cached full-map residual and the nonzero evidence measures masked reconstruction mismatch. Each utterance--ratio pair uses one seeded 16-step Euler trajectory. The detector therefore consumes a deterministic cached realization, not a Monte Carlo estimate of reconstruction uncertainty; studying repeated samples or uncertainty-aware fusion is outside the present evidence.

\paragraph{Evaluation scope.} The diagnostics are explanatory, not cross-corpus evidence. Residual and WavLM supports are fixed at 1,024 Mel frames and 4 seconds, respectively; the comparison with the single-stream reference therefore does not isolate temporal coverage. Scoring is uncalibrated, the gate is not a frequency locator, and transfer evidence is limited to ASVspoof~5-to-ITW. Missing componentwise fusion controls, matched reconstructors, whole-utterance tests, and additional targets define the claim as system level and remain priorities for follow-up work.

\section{Conclusion}

We combined multi-ratio residuals from a large-scale bona-fide DiT probe with a retained WavLM-Large branch. The primary run obtains 13.8372\% ITW EER; the three-seed mean is $15.3328\pm2.0719\%$, versus $18.2738\pm4.9847\%$ for WavLM--ResNet18. Aux raises dynamic competitive fusion from $18.4007\%$ to $25.2968\%$ in every seed. Within ASVspoof~5-to-ITW, the results support residual complementarity and a non-competitive auditory path. The strongest evidence is the all-seed interaction between auxiliary supervision and fusion structure, rather than initialization-independent superiority over the separately optimized single-stream reference. The fixed large-scale probe, different temporal supports, and system-level contrast define the present scope. Matched fusion controls, alternative reconstructors, whole-utterance inference, and additional target corpora are needed before broader causal or cross-domain claims can be made.

% ================= ACKNOWLEDGMENTS TO EDIT =================
% Uncomment and complete this block for the public preprint when applicable.
% \section*{Acknowledgments}
% Funding, institutional, and individual acknowledgments go here.
% ==========================================================

\clearpage
\bibliography{references}

@article{wang2024asvspoof5scale,
  author  = {Wang, Xin and Delgado, Hector and Tak, Hemlata and others},
  title   = {{ASVspoof 5}: Crowdsourced Speech Data, Deepfakes, and Adversarial Attacks at Scale},
  journal = {arXiv preprint arXiv:2408.08739},
  year    = {2024}
}

@article{wang2026asvspoof5evaluation,
  author  = {Wang, Xin and Delgado, Hector and Evans, Nicholas and others},
  title   = {{ASVspoof 5}: Evaluation of Spoofing, Deepfake, and Adversarial Attack Detection Using Crowdsourced Speech},
  journal = {arXiv preprint arXiv:2601.03944},
  year    = {2026}
}

@inproceedings{tak2021rawnet2,
  author    = {Tak, Hemlata and Patino, Jose and Todisco, Massimiliano and others},
  title     = {End-to-End Anti-Spoofing with {RawNet2}},
  booktitle = {Proceedings of ICASSP},
  pages     = {6369--6373},
  year      = {2021}
}

@inproceedings{jung2022aasist,
  author    = {Jung, Jee-Weon and Heo, Hee-Soo and Tak, Hemlata and others},
  title     = {{AASIST}: Audio Anti-Spoofing Using Integrated Spectro-Temporal Graph Attention Networks},
  booktitle = {Proceedings of ICASSP},
  pages     = {6367--6371},
  year      = {2022}
}

@article{chen2022wavlm,
  author  = {Chen, Sanyuan and Wang, Chengyi and Chen, Zhengyang and others},
  title   = {{WavLM}: Large-Scale Self-Supervised Pre-Training for Full Stack Speech Processing},
  journal = {IEEE Journal of Selected Topics in Signal Processing},
  volume  = {16},
  number  = {6},
  pages   = {1505--1518},
  year    = {2022}
}

@article{hsu2021hubert,
  author  = {Hsu, Wei-Ning and Bolte, Benjamin and Tsai, Yao-Hung Hubert and others},
  title   = {{HuBERT}: Self-Supervised Speech Representation Learning by Masked Prediction of Hidden Units},
  journal = {IEEE/ACM Transactions on Audio, Speech, and Language Processing},
  volume  = {29},
  pages   = {3451--3460},
  year    = {2021}
}

@inproceedings{muller2022generalize,
  author    = {M{\"u}ller, Nicolas M. and Czempin, Pavel and Dieckmann, Franziska and others},
  title     = {Does Audio Deepfake Detection Generalize?},
  booktitle = {Proceedings of Interspeech},
  pages     = {2783--2787},
  year      = {2022}
}

@article{liu2023asvspoof2021,
  author  = {Liu, Xuechen and Wang, Xin and Sahidullah, Md and others},
  title   = {{ASVspoof 2021}: Towards Spoofed and Deepfake Speech Detection in the Wild},
  journal = {IEEE/ACM Transactions on Audio, Speech, and Language Processing},
  volume  = {31},
  pages   = {2507--2522},
  year    = {2023}
}

@inproceedings{wang2024gflfad,
  author    = {Wang, Xin and Fu, Ruibo and Wen, Zhengqi and others},
  title     = {Genuine-Focused Learning Using Mask AutoEncoder for Generalized Fake Audio Detection},
  booktitle = {Proceedings of Interspeech},
  pages     = {4848--4852},
  year      = {2024}
}

@inproceedings{chen2025f5tts,
  author    = {Chen, Yushen and Niu, Zhikang and Ma, Ziyang and others},
  title     = {{F5-TTS}: A Fairytaler that Fakes Fluent and Faithful Speech with Flow Matching},
  booktitle = {Proceedings of ACL},
  pages     = {6255--6271},
  year      = {2025}
}

@inproceedings{huang2022audiomae,
  author    = {Huang, Po-Yao and Xu, Hu and Li, Juncheng and others},
  title     = {Masked Autoencoders that Listen},
  booktitle = {Advances in Neural Information Processing Systems},
  year      = {2022}
}

@inproceedings{he2022mae,
  author    = {He, Kaiming and Chen, Xinlei and Xie, Saining and others},
  title     = {Masked Autoencoders Are Scalable Vision Learners},
  booktitle = {Proceedings of CVPR},
  pages     = {16000--16009},
  year      = {2022}
}

@inproceedings{zhang2025multiview,
  author    = {Zhang, Kai and Hua, Zhen and Lan, Rushi and others},
  title     = {Multi-View Collaborative Learning Network for Speech Deepfake Detection},
  booktitle = {Proceedings of the AAAI Conference on Artificial Intelligence},
  volume    = {39},
  pages     = {1075--1083},
  year      = {2025}
}

@article{wang2024moe,
  author  = {Wang, Ziyang and Fu, Ruibo and Wen, Zhengqi and others},
  title   = {Mixture of Experts Fusion for Fake Audio Detection Using Frozen {wav2vec 2.0}},
  journal = {arXiv preprint arXiv:2409.11909},
  year    = {2024}
}

@article{zhang2021oneclass,
  author  = {Zhang, You and Jiang, Fei and Duan, Zhiyao},
  title   = {One-Class Learning Towards Synthetic Voice Spoofing Detection},
  journal = {IEEE Signal Processing Letters},
  volume  = {28},
  pages   = {937--941},
  year    = {2021}
}

@inproceedings{pascu2024generalisable,
  author    = {Pascu, Oana and Stan, Adriana and Oneata, Dan and Oneata, Elisabeta and Cucu, Horia},
  title     = {Towards Generalisable and Calibrated Audio Deepfake Detection with Self-Supervised Representations},
  booktitle = {Proceedings of Interspeech},
  pages     = {4828--4832},
  year      = {2024}
}

@inproceedings{baevski2020wav2vec2,
  author    = {Baevski, Alexei and Zhou, Yuhao and Mohamed, Abdelrahman and Auli, Michael},
  title     = {{wav2vec 2.0}: A Framework for Self-Supervised Learning of Speech Representations},
  booktitle = {Advances in Neural Information Processing Systems},
  volume    = {33},
  pages     = {12449--12460},
  year      = {2020}
}

@inproceedings{kim2021vits,
  author    = {Kim, Jaehyeon and Kong, Jungil and Son, Juhee},
  title     = {Conditional Variational Autoencoder with Adversarial Learning for End-to-End Text-to-Speech},
  booktitle = {Proceedings of ICML},
  pages     = {5530--5540},
  year      = {2021}
}

@article{kong2020diffwave,
  author  = {Kong, Zhifeng and Ping, Wei and Huang, Jiaji and Zhao, Kexin and Catanzaro, Bryan},
  title   = {{DiffWave}: A Versatile Diffusion Model for Audio Synthesis},
  journal = {arXiv preprint arXiv:2009.09761},
  year    = {2020}
}

@inproceedings{popov2021gradtts,
  author    = {Popov, Vadim and Vovk, Ivan and Gogoryan, Vladimir and Sadekova, Tasnima and Kudinov, Mikhail},
  title     = {{Grad-TTS}: A Diffusion Probabilistic Model for Text-to-Speech},
  booktitle = {Proceedings of ICML},
  pages     = {8599--8608},
  year      = {2021}
}

@inproceedings{gu2025allm4add,
  author    = {Gu, Haowen and Yi, Jiangyan and Wang, Chenglong and others},
  title     = {{ALLM4ADD}: Unlocking the Capabilities of Audio Large Language Models for Audio Deepfake Detection},
  booktitle = {Proceedings of ACM Multimedia},
  pages     = {11736--11745},
  year      = {2025}
}

@article{jung2025spoofceleb,
  author  = {Jung, Jee-Weon and Wu, Yuxuan and Wang, Xin and others},
  title   = {{SpoofCeleb}: Speech Deepfake Detection and {SASV} in the Wild},
  journal = {IEEE Open Journal of Signal Processing},
  year    = {2025},
  doi     = {10.1109/OJSP.2025.3529377}
}

@inproceedings{frank2020frequency,
  author    = {Frank, Joel and Eisenhofer, Thorsten and Sch{\"o}nherr, Lea and others},
  title     = {Leveraging Frequency Analysis for Deep Fake Image Recognition},
  booktitle = {Proceedings of ICML},
  pages     = {3247--3258},
  year      = {2020}
}

@inproceedings{liang2025trajectory,
  author    = {Liang, Yuchen and Yu, Ming and Li, Guangyuan and others},
  title     = {Denoising Trajectory Biases for Zero-Shot {AI}-Generated Image Detection},
  booktitle = {Advances in Neural Information Processing Systems},
  year      = {2025}
}

@inproceedings{li2024crossdomain,
  author    = {Li, Yiming and Zhang, Ming and Ren, Meng and others},
  title     = {Cross-Domain Audio Deepfake Detection: Dataset and Analysis},
  booktitle = {Proceedings of EMNLP},
  pages     = {4977--4983},
  year      = {2024},
  doi       = {10.18653/v1/2024.emnlp-main.286}
}

@inproceedings{zhu2024slim,
  author    = {Zhu, Yuxin and Goel, Chinmay and Koppisetti, Sree Harsha and others},
  title     = {Learn from Real: Reality Defender's Submission to {ASVspoof5} Challenge},
  booktitle = {Proceedings of the ASVspoof Workshop},
  year      = {2024},
  doi       = {10.21437/ASVspoof.2024-17}
}

@inproceedings{schafer2024robust,
  author    = {Sch{\"a}fer, Konstantin and Choi, Jee-Eun and Neu, Michael},
  title     = {Robust Audio Deepfake Detection: Exploring Front-/Back-End Combinations and Data Augmentation Strategies for the {ASVspoof5} Challenge},
  booktitle = {Proceedings of the ASVspoof Workshop},
  year      = {2024},
  doi       = {10.21437/ASVspoof.2024-9}
}

@article{tahaoglu2025nexttdnn,
  author  = {Tahaoglu, Gokhan},
  title   = {Robust DeepFake Audio Detection via an Improved {NeXt-TDNN} with Multi-Fused Self-Supervised Learning Features},
  journal = {Applied Sciences},
  volume  = {15},
  number  = {17},
  pages   = {9685},
  year    = {2025},
  doi     = {10.3390/app15179685}
}

@article{sun2024automated,
  title   = {Automated Data Augmentation for Audio Classification},
  author  = {Sun, Yanjie and Xu, Kele and Liu, Chaorun and Dou, Yong and Wang, Huaimin and Ding, Bo and Pan, Qinghua},
  journal = {IEEE/ACM Transactions on Audio, Speech, and Language Processing},
  volume  = {32},
  pages   = {2716--2728},
  year    = {2024},
  doi     = {10.1109/TASLP.2024.3402049}
}

@article{cheng2026diffrecon,
  author  = {Cheng, Bo and Cao, Songjun and Zhang, Xiaoming and Chen, Jie and Ma, Long and Chen, Fei},
  title   = {Diffusion Reconstruction towards Generalizable Audio Deepfake Detection},
  journal = {arXiv preprint arXiv:2604.26465},
  year    = {2026}
}

@inproceedings{grinberg2025diffexplain,
  author    = {Grinberg, Petr and Kumar, Ankur and Koppisetti, Surya and Bharaj, Gaurav},
  title     = {A Data-Driven Diffusion-based Approach for Audio Deepfake Explanations},
  booktitle = {Proceedings of Interspeech},
  pages     = {5348--5352},
  year      = {2025},
  doi       = {10.21437/Interspeech.2025-2105}
}

\clearpage
\onecolumn
\appendix
\setcounter{secnumdepth}{1}
\section{Additional Training and Protocol Details}\label{app:protocol}
\renewcommand{\thetable}{A\arabic{table}}
\renewcommand{\thefigure}{A\arabic{figure}}
\setcounter{table}{0}
\setcounter{figure}{0}

The overall DiT probe-training architecture is shown in Figure~\ref{fig:dit-training} in the main text and is not repeated here. This section consolidates the preprocessing, cache-generation, detector-training, model-selection, and scoring protocols used by the residual-based systems and the separately optimized single-stream reference. The DiT and the frozen Audio-MAE condition encoder use only bona fide speech; no spoof label or attack identity is provided to the reconstruction probe.

The exact DiT training-corpus composition and detector split statistics are reported in Tables~\ref{tab:dit-training-data} and~\ref{tab:detector-data} in the main text. Table~\ref{tab:frontend-supp} summarizes the paper-level settings needed to interpret the controlled comparisons without duplicating the method description in the main text.

\begin{table}[H]
\centering
\caption{Input support, residual generation, and final scoring protocols. Baseline-specific checkpoint selection is summarized below.}
\label{tab:frontend-supp}
\scriptsize
\begin{tabularx}{\columnwidth}{L{0.31\columnwidth} X}
\toprule
Item & Setting \\
\midrule
Waveform sampling & 16 kHz \\
Spectrogram--residual support & Leading 1,024 Mel frames ($\approx$10.24 s at 160-sample hop); feature-domain padding when shorter \\
Residual systems: WavLM train support & Random 64,000-sample (4 s) waveform crop; tail zero-padding when shorter \\
Residual systems: Dev/Eval/ITW/prototype & Deterministic leading 64,000 samples (4 s); tail zero-padding \\
WavLM--ResNet18 train/Dev support & Random 64,000-sample (4 s) crop during training and Dev checkpoint selection \\
WavLM--ResNet18 final scoring support & Leading 64,000 samples for Train prototype, Eval, and ITW; tail zero-padding \\
Spectral representation & 128-bin Mel spectrogram; one shared cache-generation pipeline for all residual-based systems \\
Residual definition & $R^{(r)}=|X-\widehat X^{(r)}|$; visible cells are copied from $X$, so residuals are nonzero only on masked cells; no sample-wise input InstanceNorm \\
Mask geometry & Independent Bernoulli masking over time--frequency cells with expected masked fraction $r$ \\
Masking ratios / solver steps & $\{0.5,0.75,0.9\}$; uniform-grid Euler integration with 16 steps per ratio \\
Detector input & Original Mel plus three absolute residual maps \\
Inference score & Negative cosine similarity to the ASVspoof 5 Train bona fide prototype \\
Target-domain adaptation & None: no center update, threshold fitting, or score calibration \\
\bottomrule
\end{tabularx}
\end{table}

The reconstruction probe uses a frozen Audio-MAE-Base encoder (patch size 16, width 768, depth 12, 12 heads; decoder removed) to condition a DiT of width 1,024, depth 22, 16 attention heads, and feed-forward multiplier 2. During flow matching and sampling, visible time--frequency cells are clamped to the original Mel values and assigned zero velocity. One stochastic reconstruction is generated per utterance and ratio under fixed export seeds and classifier-free guidance strength 1.0, then cached; all residual-based systems, seeds, splits, and ablations reuse these files. Thus, Equation (4) is evaluated over the full tensor but has zero visible-region error by construction. The settings below specify the frontend, model, masking, and export protocol used for the reported residual caches.

For the four residual-based configurations, detector training uses five epochs of AdamW with initial learning rate $10^{-3}$, weight decay $10^{-4}$, one warm-up epoch, and cosine decay. Physical batches contain 32 bona fide and 32 spoofed samples balanced across A01--A08; gradients accumulate for four physical batches, while batch-hard mining remains within each batch. The triplet margin is 0.3 and the embedding dimension is 256. The 33-way source head uses label smoothing 0.1 and weight 0.2 when enabled. The fixed output index covers bona fide plus A01--A32, but positive training targets occur only for bona fide and A01--A08; A09--A32 are reserved indices and never enter training as observed identities. This is an indexing convention, not a claimed open-set objective. Each masking ratio uses 16 DiT reconstruction steps.

The WavLM--ResNet18 reference is separately optimized to provide a strong auditory-only baseline. Frozen WavLM-Large hidden states are learnably aggregated across 25 outputs, rearranged from $[B,T,1024]$ to a single-channel $[B,1,1024,T]$ map, and encoded by a randomly initialized ResNet-18 into 256 dimensions. It uses five epochs, the same triplet margin and source auxiliary objective, gradient accumulation of four, and seeds 42, 123, and 2026, but uses learning rate $10^{-4}$ and a binary class-balanced WeightedRandomSampler rather than attack-balanced physical batches.

All six reported configurations are independently trained with seeds 42, 123, and 2026, but their roles and Dev protocols differ. The four residual-based configurations select the lowest-Dev-EER checkpoint using deterministic leading waveform crops and a bona fide prototype from ASVspoof 5 Train. The two WavLM--ResNet18 configurations follow their original baseline procedure: Dev uses random 4-second crops and a Dev-derived bona fide center for checkpoint selection. After selection, every final Eval and ITW result in this study recomputes the bona fide prototype from ASVspoof 5 Train and uses deterministic leading 4-second crops. Eval and ITW never participate in checkpoint selection, prototype adaptation, or threshold calibration. Seed 42 was fixed before repeated-seed validation; results use mean and sample standard deviation over three seeds.

The prototype score is used as an uncalibrated ranking score. We report EER and official normalized min-DCF, but do not claim calibrated likelihoods or operational actDCF performance. The present cross-corpus claim is restricted to ASVspoof 5-to-ITW transfer. The residual systems share data, cache, auditory branch, objectives, optimizer, and scoring, but the dynamic comparison also changes residual routing and visual organization; it is therefore a system-level contrast rather than a componentwise matched anchoring ablation. Additional multilingual, platform, codec-specific, whole-utterance, matched-fusion, and alternative-reconstructor evaluations remain outside the current experiment set.

\begin{table}[H]
\centering
\caption{Offline reconstruction and cached detector cost per utterance.}
\label{tab:cost-supp}
\scriptsize
\begin{tabularx}{\columnwidth}{L{0.22\columnwidth} X C{0.16\columnwidth} L{0.18\columnwidth}}
\toprule
Item & Setting & Cost & Note \\
\midrule
Multi-ratio residual generation & Three ratios, 16 DiT steps each & 2.37 s & Main offline cost \\
Cached dual-stream forward pass & Spectrogram residuals + WavLM & 24.4 ms & Excludes disk I/O \\
Residual cache & NPZ multi-ratio features & $\approx$1.00 MiB & Linear in corpus size \\
\bottomrule
\end{tabularx}
\end{table}

\FloatBarrier
\section{Three-Seed Validation}\label{app:three-seed}
\renewcommand{\thetable}{B\arabic{table}}
\renewcommand{\thefigure}{B\arabic{figure}}
\setcounter{table}{0}
\setcounter{figure}{0}

\begin{table}[H]
\centering
\caption{Three-seed summary. WavLM--ResNet18 is a separately optimized strong single-stream reference; the four residual-based rows form the controlled fusion comparison.}
\label{tab:three-seed-summary}
\tiny
\setlength{\tabcolsep}{2.2pt}
\renewcommand{\arraystretch}{0.96}
\resizebox{\textwidth}{!}{%
\begin{tabular}{lcccccc}
\toprule
System & Aux. & Dev EER & Eval EER & Eval min-DCF & ITW EER & ITW min-DCF \\
\midrule
WavLM--ResNet18 & No & $4.8707\pm0.4186$ & $6.9306\pm0.3420$ & $0.18952\pm0.01937$ & $22.5704\pm4.7675$ & $0.54558\pm0.07409$ \\
WavLM--ResNet18 & Yes & $4.9445\pm0.2190$ & $7.0603\pm0.5564$ & $0.19076\pm0.01520$ & $18.2738\pm4.9847$ & $0.42791\pm0.08148$ \\
Audio-anchored & No & $2.4100\pm1.0732$ & $7.1112\pm0.3727$ & $0.19465\pm0.00662$ & $17.1328\pm5.8921$ & $0.45984\pm0.14040$ \\
Audio-anchored & Yes & $1.7595\pm0.3997$ & $6.8885\pm0.3308$ & $0.19007\pm0.00498$ & $\mathbf{15.3328\pm2.0719}$ & $\mathbf{0.41776\pm0.06263}$ \\
Dynamic competitive & No & $2.9382\pm0.3183$ & $6.7901\pm0.8929$ & $0.18528\pm0.02248$ & $18.4007\pm2.5810$ & $0.43718\pm0.04155$ \\
Dynamic competitive & Yes & $1.5041\pm0.3323$ & $7.1585\pm0.2819$ & $0.20216\pm0.00681$ & $25.2968\pm1.8357$ & $0.60764\pm0.08569$ \\
\bottomrule
\end{tabular}%
}
\end{table}

\begin{table}[H]
\centering
\caption{Complete per-seed results. Final Eval and ITW use a separately computed ASVspoof 5 Train bona fide prototype; checkpoint selection follows the residual-system or WavLM--ResNet18 Dev protocol described above.}
\label{tab:three-seed-full}
\tiny
\setlength{\tabcolsep}{2.5pt}
\renewcommand{\arraystretch}{0.93}
\begin{tabularx}{\textwidth}{L{0.27\textwidth} C{0.05\textwidth} C{0.06\textwidth} C{0.10\textwidth} C{0.11\textwidth} C{0.11\textwidth} C{0.10\textwidth} C{0.11\textwidth}}
\toprule
System & Aux. & Seed & Dev EER & Eval EER & Eval min-DCF & ITW EER & ITW min-DCF \\
\midrule
WavLM--ResNet18 single-stream & No & 42 & 4.3875 & 7.2699 & 0.20943 & 27.1452 & 0.61732 \\
WavLM--ResNet18 single-stream & No & 123 & 5.1034 & 6.5860 & 0.17074 & 22.9350 & 0.55006 \\
WavLM--ResNet18 single-stream & No & 2026 & 5.1213 & 6.9358 & 0.18839 & 17.6311 & 0.46935 \\
WavLM--ResNet18 single-stream & Yes & 42 & 5.1472 & 6.5226 & 0.17533 & 18.5994 & 0.40741 \\
WavLM--ResNet18 single-stream & Yes & 123 & 4.7122 & 7.0247 & 0.19123 & 23.0877 & 0.51768 \\
WavLM--ResNet18 single-stream & Yes & 2026 & 4.9741 & 7.6337 & 0.20571 & 13.1343 & 0.35864 \\
Audio-anchored additive fusion & No & 42 & 1.8159 & 6.6809 & 0.18702 & 19.2282 & 0.50756 \\
Audio-anchored additive fusion & No & 123 & 3.6488 & 7.3348 & 0.19808 & 10.4794 & 0.30180 \\
Audio-anchored additive fusion & No & 2026 & 1.7652 & 7.3179 & 0.19886 & 21.6909 & 0.57017 \\
Audio-anchored additive fusion & Yes & 42 & 1.8586 & 6.5442 & 0.18456 & 13.8372 & 0.36921 \\
Audio-anchored additive fusion & Yes & 123 & 2.1004 & 7.2039 & 0.19424 & 14.4634 & 0.39562 \\
Audio-anchored additive fusion & Yes & 2026 & 1.3196 & 6.9174 & 0.19140 & 17.6977 & 0.48846 \\
Dynamic residual routing with competitive fusion & No & 42 & 3.2766 & 6.0208 & 0.16414 & 20.9337 & 0.47368 \\
Dynamic residual routing with competitive fusion & No & 123 & 2.8931 & 6.5802 & 0.18280 & 18.4942 & 0.44589 \\
Dynamic residual routing with competitive fusion & No & 2026 & 2.6449 & 7.7692 & 0.20889 & 15.7742 & 0.39197 \\
Dynamic residual routing with competitive fusion & Yes & 42 & 1.1505 & 6.8333 & 0.19430 & 23.2029 & 0.50870 \\
Dynamic residual routing with competitive fusion & Yes & 123 & 1.8099 & 7.3316 & 0.20583 & 26.6293 & 0.65608 \\
Dynamic residual routing with competitive fusion & Yes & 2026 & 1.5518 & 7.3107 & 0.20634 & 26.0582 & 0.65814 \\
\bottomrule
\end{tabularx}
\end{table}

The WavLM--ResNet18 rows position the complete detector against a strong, separately optimized auditory-only reference; they are not one-branch ablations. Paired by seed, audio anchoring with auxiliary supervision improves over this reference for seeds 42 and 123 but not 2026, so we claim a lower mean rather than initialization-independent superiority. Against dynamic competitive fusion with auxiliary supervision, the audio-anchored system improves ITW EER in all three seeds by 9.3657, 12.1659, and 8.3605 points. Auxiliary supervision itself changes the dynamic system by +2.2692, +8.1351, and +10.2840 points. With only three seeds, no statistical-significance claim is made. These repeated outcomes support a fusion--supervision interaction, while the routing and visual-encoder differences prevent a componentwise causal attribution to anchoring alone.

\FloatBarrier
\section{Additional Ablation Results}\label{app:ablation}
\renewcommand{\thetable}{C\arabic{table}}
\renewcommand{\thefigure}{C\arabic{figure}}
\setcounter{table}{0}
\setcounter{figure}{0}

\begin{table}[H]
\centering
\caption{Ablation of the frequency squeeze-and-excitation (Freq-SE) module in the audio-anchored complete model.}
\label{tab:freqse-supp}
\scriptsize
\begin{tabular}{lccc}
\toprule
Setting & Dev EER (\%) & Eval EER (\%) & Eval min-DCF \\
\midrule
Without Freq-SE & 1.4739 & 6.6042 & 0.18281 \\
Complete model & 1.8586 & 6.5442 & 0.18456 \\
\bottomrule
\end{tabular}
\end{table}

Removing Freq-SE increases Eval EER from 6.5442\% to 6.6042\%, while Dev EER and Eval min-DCF improve slightly. Because the differences are small and metric directions are not fully aligned, Freq-SE is treated as a lightweight frequency-modeling component rather than a principal performance source.

The main text verifies each individual masking ratio and their three-ratio combination. It does not exhaustively enumerate every two-ratio subset or a denser masking grid, and 16 reconstruction steps are a fixed operating point rather than a claimed optimum. These finer performance--cost trade-offs are left for future work.

\FloatBarrier
\section{Extended Mask-Ratio Diagnostics}\label{app:diagnostics}
\renewcommand{\thetable}{D\arabic{table}}
\renewcommand{\thefigure}{D\arabic{figure}}
\setcounter{table}{0}
\setcounter{figure}{0}

\begin{figure}[H]
    \centering
    \begin{minipage}[t]{0.49\textwidth}
        \centering
        \includegraphics[width=\linewidth,height=0.245\textheight,keepaspectratio]{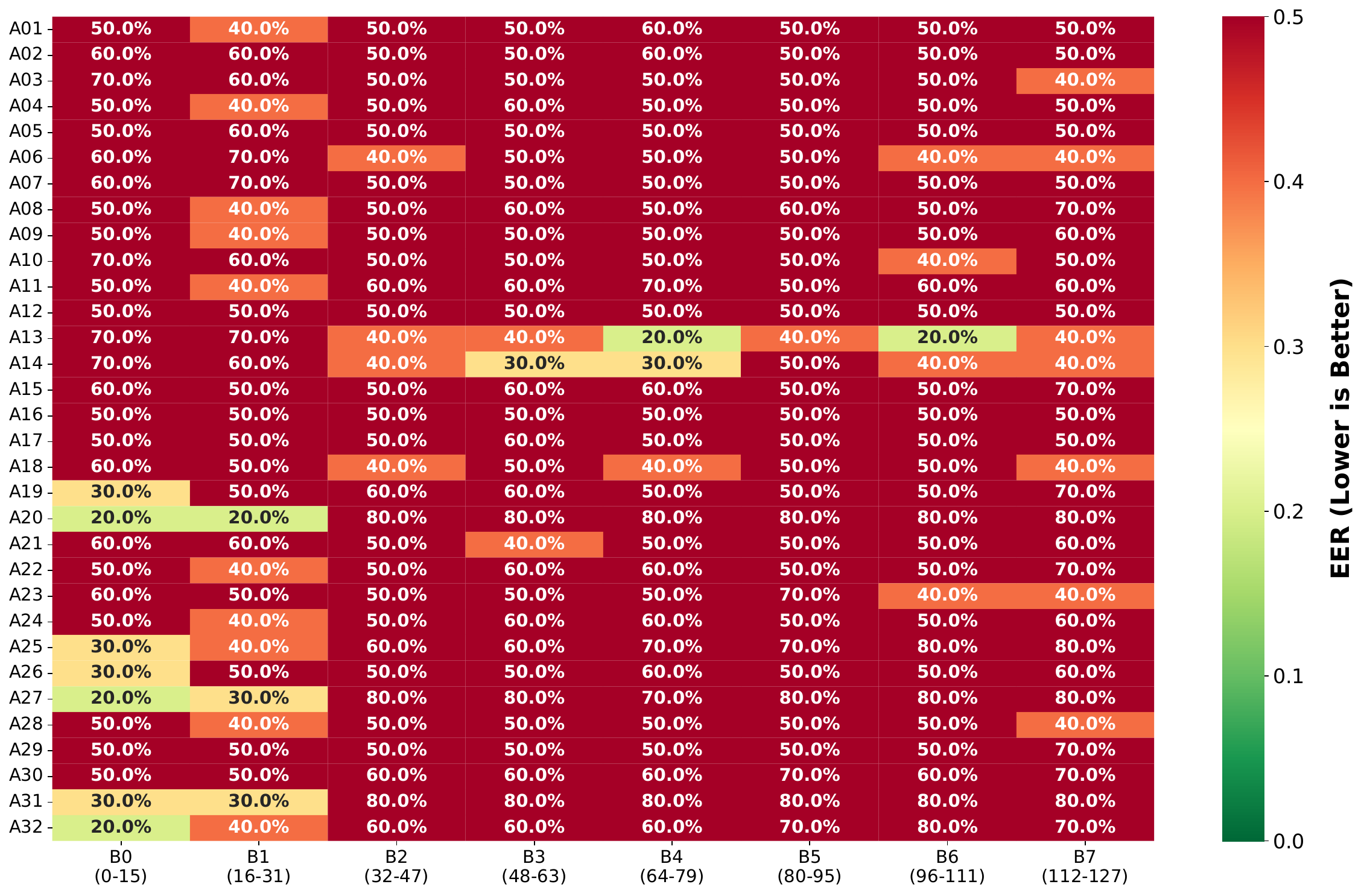}\\[-0.4ex]
        {\small (a) Per-attack subband EER, ratio 0.1}
    \end{minipage}
    \hfill
    \begin{minipage}[t]{0.49\textwidth}
        \centering
        \includegraphics[width=\linewidth,height=0.245\textheight,keepaspectratio]{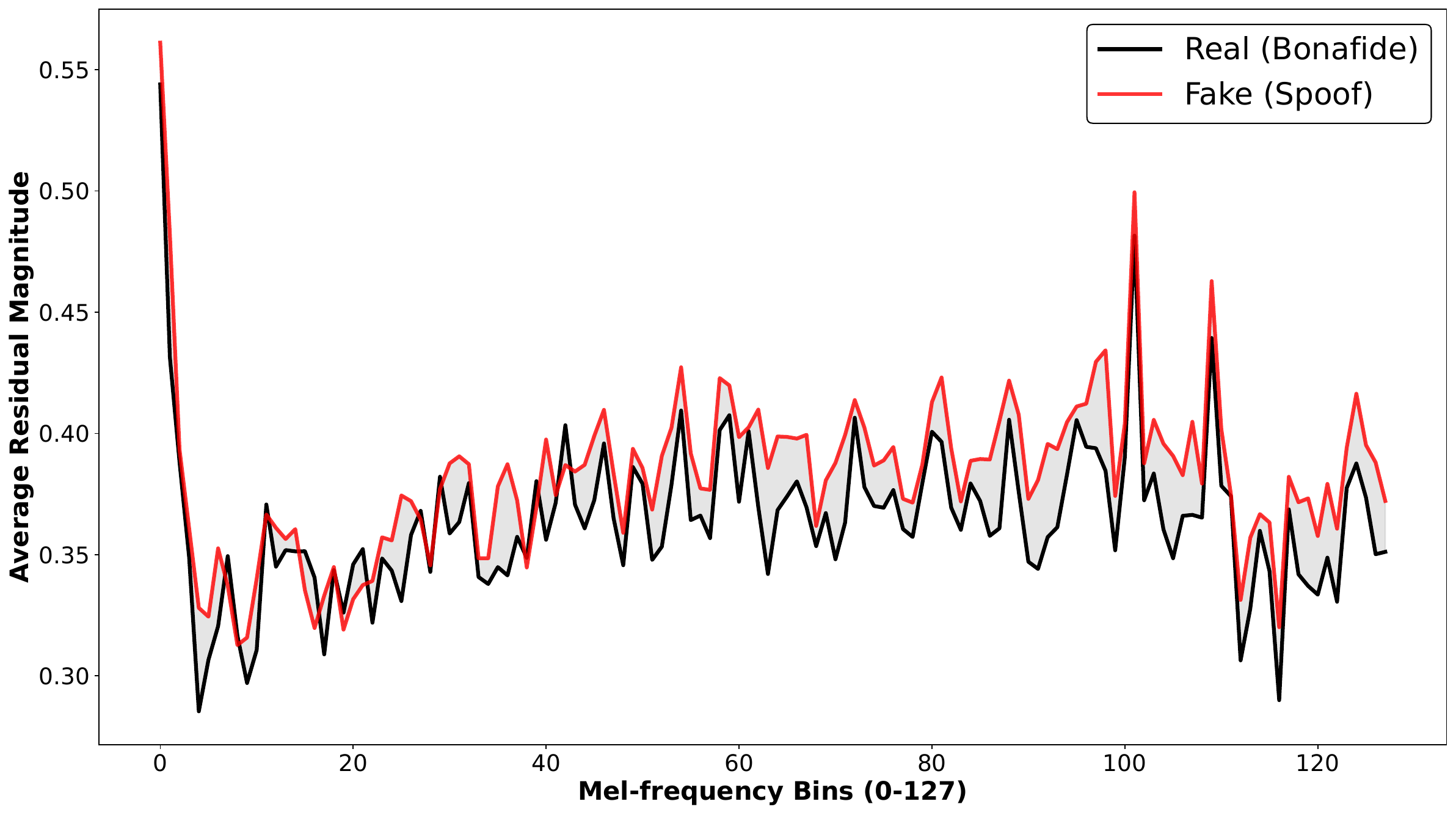}\\[-0.4ex]
        {\small (b) Mean residual spectrum, ratio 0.1}
    \end{minipage}

    \vspace{1.5mm}

    \begin{minipage}[t]{0.49\textwidth}
        \centering
        \includegraphics[width=\linewidth,height=0.245\textheight,keepaspectratio]{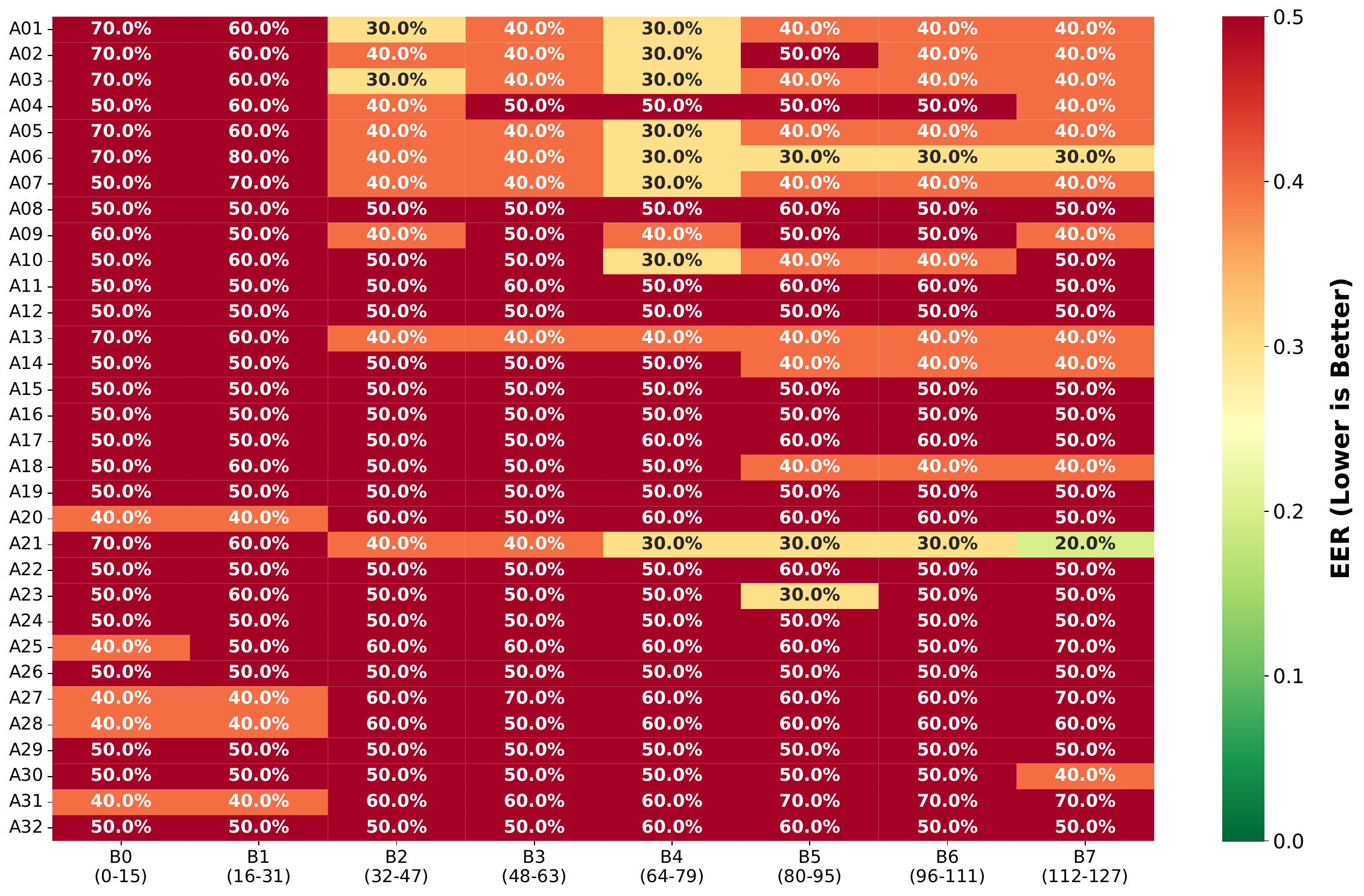}\\[-0.4ex]
        {\small (c) Per-attack subband EER, ratio 0.3}
    \end{minipage}
    \hfill
    \begin{minipage}[t]{0.49\textwidth}
        \centering
        \includegraphics[width=\linewidth,height=0.245\textheight,keepaspectratio]{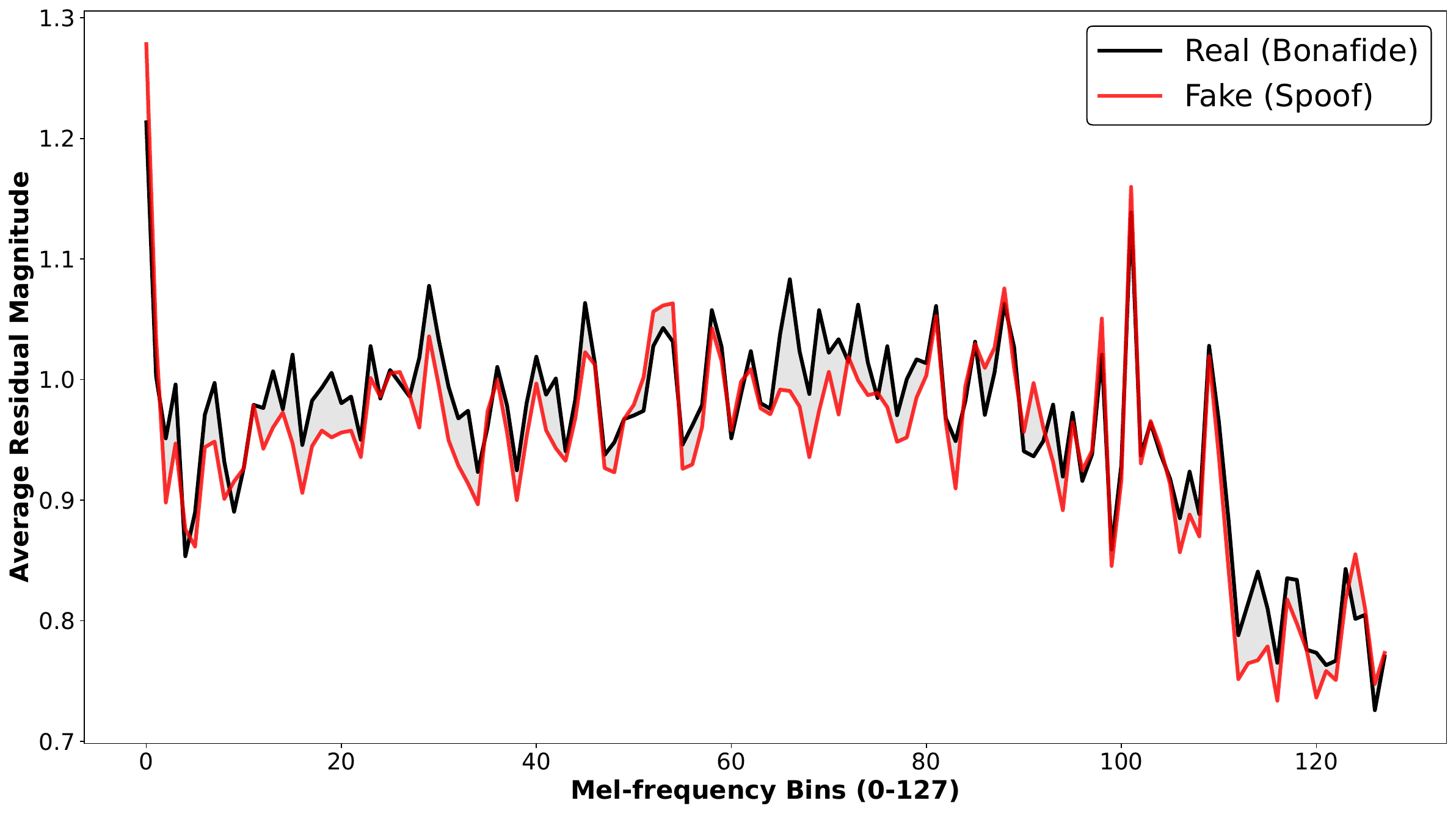}\\[-0.4ex]
        {\small (d) Mean residual spectrum, ratio 0.3}
    \end{minipage}
    \caption{Extended diagnostics at low masking ratios. With abundant visible context, the residual-spectrum separation is comparatively broad rather than concentrated in a universal subband, while the per-attack heatmaps already show substantial heterogeneity. These plots are explanatory and are not detector-performance comparisons.}
    \label{fig:maskdiag-low}
\end{figure}

\begin{figure}[H]
    \centering
    \begin{minipage}[t]{0.49\textwidth}
        \centering
        \includegraphics[width=\linewidth,height=0.245\textheight,keepaspectratio]{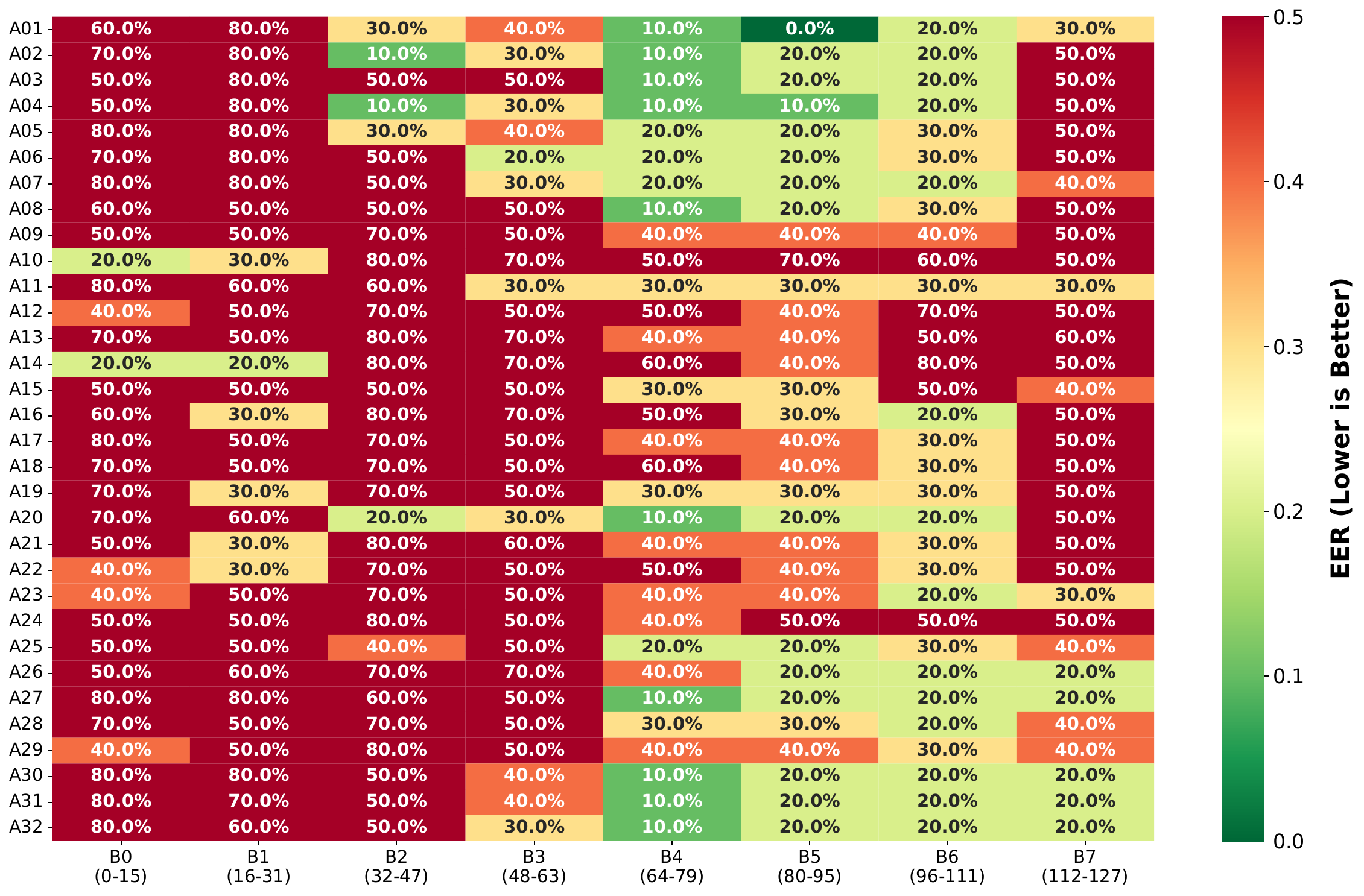}\\[-0.4ex]
        {\small (a) Per-attack subband EER, ratio 0.5}
    \end{minipage}
    \hfill
    \begin{minipage}[t]{0.49\textwidth}
        \centering
        \includegraphics[width=\linewidth,height=0.245\textheight,keepaspectratio]{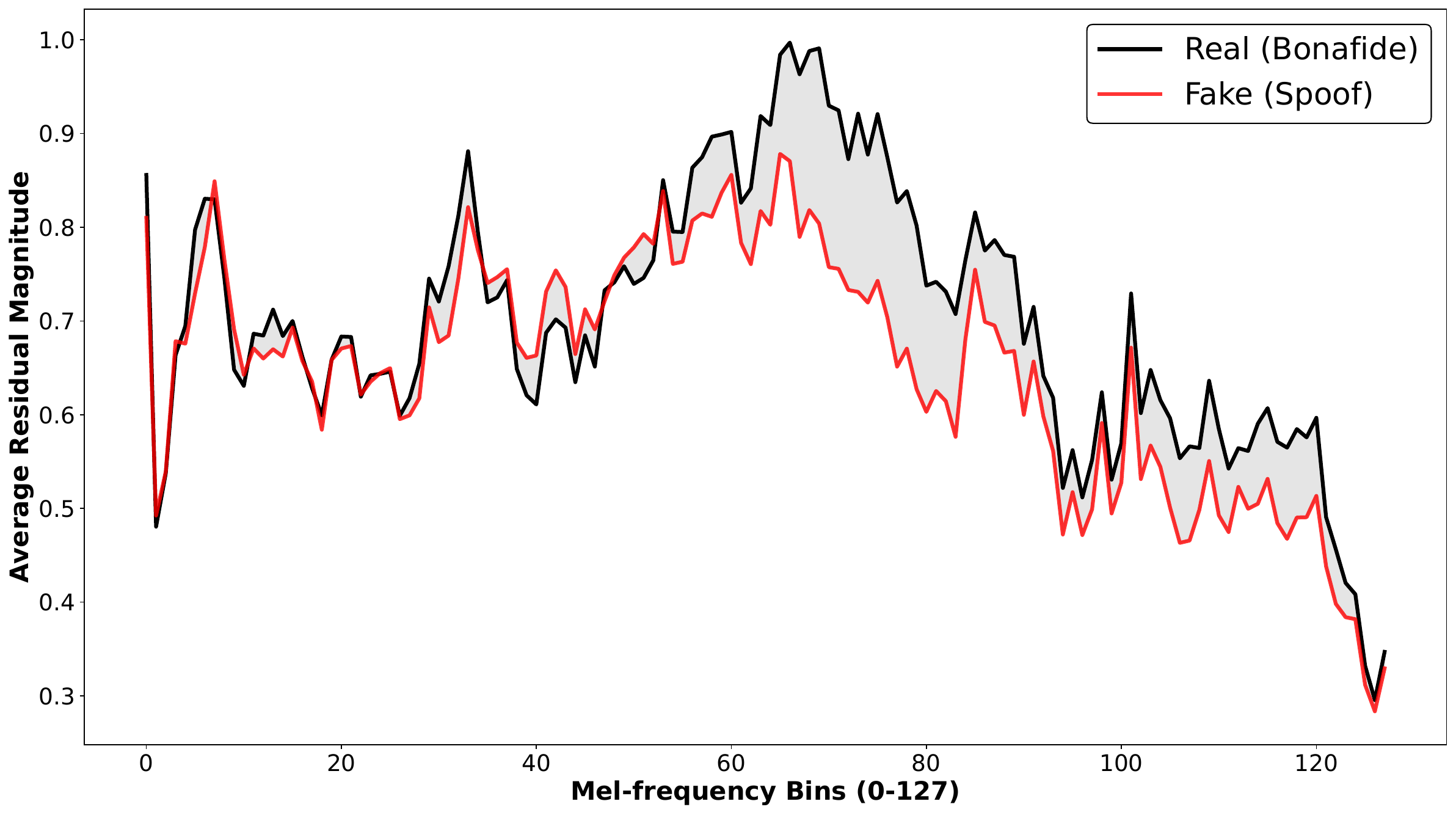}\\[-0.4ex]
        {\small (b) Mean residual spectrum, ratio 0.5}
    \end{minipage}

    \vspace{1.5mm}

    \begin{minipage}[t]{0.49\textwidth}
        \centering
        \includegraphics[width=\linewidth,height=0.245\textheight,keepaspectratio]{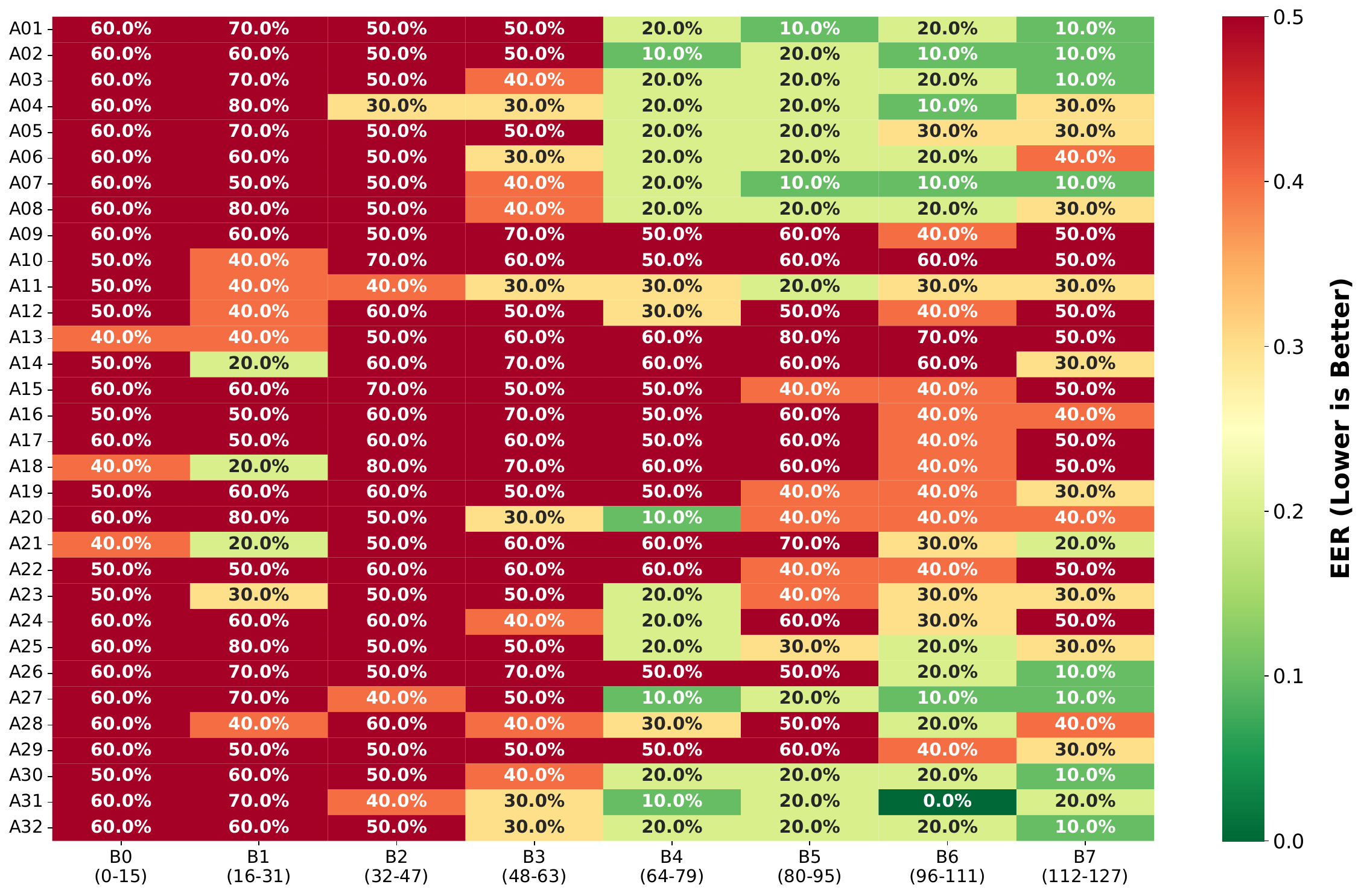}\\[-0.4ex]
        {\small (c) Per-attack subband EER, ratio 0.75}
    \end{minipage}
    \hfill
    \begin{minipage}[t]{0.49\textwidth}
        \centering
        \includegraphics[width=\linewidth,height=0.245\textheight,keepaspectratio]{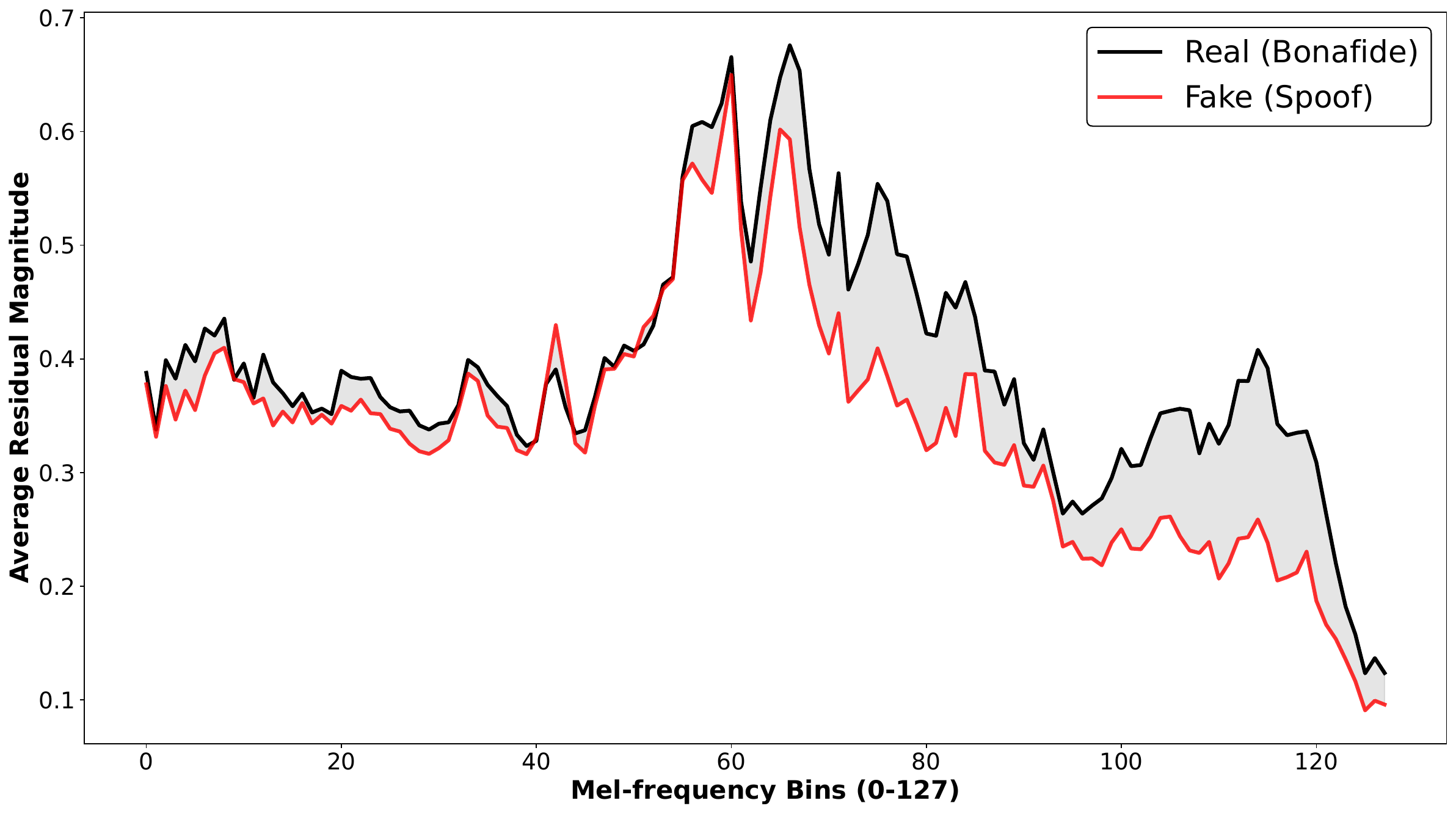}\\[-0.4ex]
        {\small (d) Mean residual spectrum, ratio 0.75}
    \end{minipage}
    \caption{Extended diagnostics at the masking ratios used by, or close to, the detector. As masking difficulty increases, both the relative bona-fide/spoof residual trends and the most discriminative subbands change. Together with the ratio-0.9 diagnostics in the main text, these results motivate retaining multiple reconstruction difficulties, but they do not imply a universal frequency boundary or a monotonic relation between masking ratio and detection performance.}
    \label{fig:maskdiag-mid}
\end{figure}

The main text reports the residual-spectrum diagnostic and per-attack subband heatmap at masking ratio 0.9. Both use the balanced diagnostic probe manifest spanning A01--A32; the exporter averages the absolute residual over time for each utterance and the plotting script then averages over samples, without detector training or sample-wise normalization. To provide a broader view of how the reconstruction evidence changes with masking difficulty, Figures~\ref{fig:maskdiag-low} and~\ref{fig:maskdiag-mid} show the corresponding post-hoc diagnostics at ratios 0.1, 0.3, 0.5, and 0.75. Ratios 0.1 and 0.3 are included only for visualization and are not inputs to the detector. For every ratio, the heatmap reports the EER obtained by using the mean absolute residual in each of eight Mel subbands as a scalar score for each attack A01--A32, while the spectrum plot averages the residual over time and then over samples. These analyses were performed after the model design was fixed and were not used for ratio selection, training, checkpoint selection, prototype construction, or threshold calibration.

The diagnostic sweep complements, but does not replace, the controlled detector ablation in the main text. The latter shows that the joint input at ratios 0.5, 0.75, and 0.9 outperforms every corresponding single-ratio detector on ASVspoof 5 Eval. The visualizations here explain why complementarity is plausible: changing the reconstruction difficulty alters both the global residual profile and the attack-specific subband ranking. Because the plots are computed post hoc and include Eval attacks, they are interpreted only as descriptive evidence.

\FloatBarrier

\end{document}